\documentclass[twocolumn]{aastex701}

\begin{document}
\title{Asteroseismic analysis of RY Leporis: the post-main sequence HADS in a binary system}

\author[orcid=0009-0001-8106-9186,gname=Wojciech,sname=Niewiadomski]{Wojciech Niewiadomski}
\affiliation{University of Wroclaw, Faculty of Physics and Astronomy, Astronomical Institute, ul. Kopernika 11, PL-51-622 Wroc\l{}aw, Poland}
\email[show]{wojciech.niewiadomski@uwr.edu.pl}  

\author[orcid=0000-0001-9704-6408,gname=Jadwiga, sname=Daszy{\'n}ska-Daszkiewicz]{Jadwiga Daszy{\'n}ska-Daszkiewicz}
\affiliation{University of Wroclaw, Faculty of Physics and Astronomy, Astronomical Institute, ul. Kopernika 11, PL-51-622 Wroc\l{}aw, Poland}
\email{daszynska@astro.uni.wroc.pl}

\author[orcid=0000-0003-3476-8483,gname=Przemys\l{}aw,sname=Walczak]{Przemys\l{}aw Walczak}
\affiliation{University of Wroclaw, Faculty of Physics and Astronomy, Astronomical Institute, ul. Kopernika 11, PL-51-622 Wroc\l{}aw, Poland}
\email{walczak@astro.uni.wroc.pl}

\author[orcid=0000-0002-2393-8427,gname=Wojciech,sname=Szewczuk]{Wojciech Szewczuk}
\affiliation{University of Wroclaw, Faculty of Physics and Astronomy, Astronomical Institute, ul. Kopernika 11, PL-51-622 Wroc\l{}aw, Poland}
\email{szewczuk@astro.uni.wroc.pl}

\author[orcid=0000-0001-6827-9077,gname=Eloy,sname=Rodr\'iguez]{Eloy Rodr\'iguez}
\affiliation{Instituto de Astrof\'isica de Andaluc\'ia, CSIC, PO Box 3004, E-18080 Granada, Spain}

\email{eloy@iaa.csic.es}

\author[orcid=0000-0003-2244-1512,gname=Piotr Antoni,sname=Ko\l{}aczek-Szyma{\'n}ski]{Piotr Ko\l{}aczek-Szyma{\'n}ski}
\affiliation{University of Wroclaw, Faculty of Physics and Astronomy, Astronomical Institute, ul. Kopernika 11, PL-51-622 Wroc\l{}aw, Poland}
\affiliation{STAR Institute, University of Li\`ege, 19C All\'ee du 6 Ao\^ut, B-4000 Li\`ege, Belgium}
\email{piotr.kolaczek-szymanski@uwr.edu.pl}

\author[orcid=0000-0002-6526-9444,gname=Aliz,sname=Derekas]{Aliz Derekas}
\affiliation{ELTE E\"otv\"os L\`or\`and University, Gothard Astrophysical Observatory, Szombathely Szent Imre h. u. 112, H-9700, Hungary}
\email{karton67@gmail.com}

\begin{abstract}
We present a comprehensive study of the pulsating primary component of the long-period binary system RY~Lep. The spectral energy distribution and the absence of detectable lines indicate that the companion is likely a white dwarf. Atmospheric parameters and chemical abundances were determined from a high-resolution spectrum obtained with the the Southern African Large Telescope (SALT). The spectroscopic analysis reveals 
an underabundance of iron-group elements and an enhancement of neutron-capture elements, including barium and europium, with an overall metallicity of [m/H]$\approx -0.4$. 
In the next step, we performed the first Fourier analysis of long-term photometric data from ASAS, SuperWASP, and TESS. In the TESS observations, we identify several additional frequencies not present in the ground-based data. The dominant frequency at 4.4415\,d$^{-1}$ is identified as a radial mode, most likely the first radial overtone.
Finally, seismic modeling of RY~Lep was carried out by fitting the dominant mode together with the secondary frequency at 6.5991\,d$^{-1}$,
considering two possible identification for the latter: the third radial overtone or a dipole mode. Bayesian inference based on Monte Carlo simulations yields a stellar mass of $\sim 2.0\,$M$_\odot$ and an age of $\sim 730$\,Myr. All viable seismic models place RY~Lep in the hydrogen shell–burning evolutionary phase, with a metallicity consistent with the spectroscopic determination.
\end{abstract}

\section{Introduction}
\label{sec:introduction}
Asteroseismology provides precise constraints on the global parameters, internal structure, and evolutionary states of stars. Among the most extensively studied classes of pulsating variables are the $\delta$~Scuti stars—short-period pulsators with masses of approximately $1.6$–$2.6$\,M$_\odot$ and spectral types A–F. They are Population~I objects found predominantly on the main sequence \citep{1998A&A...332..958B, 2016MNRAS.460.1970B}. Similar to RR~Lyrae and classical Cepheid variables, pulsations in $\delta$~Sct stars are driven by the $\kappa$ mechanism operating in the second helium-ionization zone \citep{1971A&A....14...24C}.

$\delta$~Sct stars with securely identified pulsation modes play an important role in testing models of stellar structure and evolution.
A particularly important subclass of these pulsators is the high-amplitude $\delta$~Scuti (HADS) stars, which exhibit brightness variations in the $V$ band exceeding 0.3\,mag \citep{2000ASPC..210....3B}. Their pulsations are typically dominated by one or two radial modes, sometimes accompanied by additional low-amplitude frequencies. HADS stars are generally slow rotators, with typical projected rotational velocities ($V_{\rm rot}\sin i$) below 40\,km\,s$^{-1}$ \citep{2000ASPC..210....3B}. 
The presence of two radial modes imposes particularly strong constraints on stellar models because their frequency ratios are confined to a narrow range. The commonly observed mode pair corresponds to the fundamental and first-overtone radial modes, with a typical frequency ratio of $\sim 0.77$.

Seismic studies of double-mode, radially pulsating HADS stars have shown that these objects are usually in a post–main-sequence evolutionary stage, burning hydrogen in a shell \citep[e.g.,][]{1996A&A...312..463P, 2001A&A...366..178R, 2023ApJ...942L..38D}. Recent work has also demonstrated that seismic models of double-mode $\delta$~Sct stars are highly sensitive to the adopted opacity data. In most cases, only models computed with OPAL opacities reproduce the frequencies of the fundamental and first radial overtone while simultaneously matching the observed $(T_{\rm eff},L)$ values \citep{2023ApJ...942L..38D}. Models based on OP \citep{2005MNRAS.362L...1S} or OPLIB \citep{2016ApJ...817..116C} opacity tables tend to be significantly cooler and less luminous.

The radial nature of HADS pulsations can also be verified through multicolor photometry or time-resolved spectroscopy. Multicolor photometry further enables the determination of empirical values of the nonadiabatic parameter $f$, which quantifies the ratio of radiative-flux variation to radial displacement at the photosphere. Comparing empirical $f$ values with those from linear nonadiabatic pulsation calculations provides a powerful diagnostic of, for example, the efficiency of convective energy transport. Detailed seismic modeling, yielding stringent constraints on opacity data and envelope convection, have been carried out for, e.g., SX Phe \citep{2020MNRAS.499.3034D}, BP Peg \citep{2022MNRAS.512.3551D}, 
AE UMa and RV Ari \citep{2023MNRAS.526.1951D}.

In this paper, we present a comprehensive analysis of the pulsating primary component of the RY~Lep system. Section~\ref{sec:RY_Lep_description} summarizes basic information about the star. In Section~\ref{sec:RY_Lep_photometry}, we review atmospheric parameters derived from photometry, including both literature results and our estimates based on model atmospheres. Section~\ref{sec:RY_Lep_spectroscopy} presents atmospheric parameters and chemical abundances obtained from SALT and SSO spectra. The Fourier analysis of TESS, ASAS, and SuperWASP light curves is described in Section~\ref{sec:Fourier_analysis}. In Section~\ref{sec:mode_identification}, we identify the dominant pulsation mode, and Section~\ref{sec:Seismic_modeling} details the seismic modeling of RY~Lep using a Monte Carlo–based Bayesian approach. Section~\ref{sec:discussion} discusses the implications of our findings, and Section~\ref{sec:summary} summarizes the main results.

\section{RY Lep}\label{sec:RY_Lep_description}
RY~Leporis (HD~38882, RY~Lep) is a star with a mean brightness in the $V$ passband of $8.25$\,mag \citep{2000A&A...355L..27H}. Its first spectral classification by \citet{1966AJ.....71S.175P} was F0, whereas \citet{1988mcts.book.....H} assigned it as A9V.

The variability of RY~Lep was first discovered by \citet{1964IBVS...51....1S}, who initially classified it as an Algol-type binary based on two recorded times of minimum light. \citet{1985A&A...149..465D} reclassified RY~Lep as a relatively bright high-amplitude $\delta$~Sct star, with a dominant pulsation period of $P = 0.2254$\,d and a brightness range in the Johnson $V$ passband of $\Delta V = 0.35$\,mag. The dominant frequency, $4.4415$\,d$^{-1}$, has a relatively stable amplitude and is observed in both light and radial velocity variations \citep{2004CoAst.145...48R}. The radial nature of this pulsation was suggested by \citet{1995MNRAS.277..965R} from phase differences in the $uvby$ and $BV$ filters. \citet{2004CoAst.145...48R} also reported a second frequency of approximately $6.6$\,d$^{-1}$, which exhibited strongly modulated amplitude. From phase differences among the $uvby$ light curves, they suggested that this frequency could correspond to a nonradial mode. The $6.6$\,d$^{-1}$ frequency was later confirmed by \citet{2009MNRAS.394..995D}, who proposed it could be radial, as the frequency ratio of $4.4415$\,d$^{-1}$ and $6.6$\,d$^{-1}$ is consistent with the first and third overtone radial modes.

Quite recently, \citet{2023A&A...674A..18C} classified RY~Lep as an RR~Lyrae-type variable based on its Gaia $G$-band amplitude ($\Delta G > 0.1$\,mag) and period ($P \in [0.2, 1.0]$\,d). From its location on the period–amplitude diagram, they further identified it as an RRc star. It will be discussed later that such a classification is not reliable.

There are also indications of the binary nature of RY~Lep. The first mention appears in \citet{2002ASPC..256..173L,2002ASPC..259..112L}, based on radial velocity measurements and the light-travel time effect
in $J$-band photometry. The orbital period was estimated to be about 730\,d \citep{2003ASPC..292..203L}. Long-term radial
velocity variability was later confirmed by \citet{2009MNRAS.394..995D} from a nearly 700\,d dataset. They estimated the mass 
of the secondary to be $1.10(15)$\,M$_{\sun}$ and the orbital semi-major axis to be 2.3\,AU, assuming an orbital period of 730\,d and a primary mass of $M_1 = 2.16(10)$\,M$_{\sun}$ \citep{1995MNRAS.277..965R}. The absence of spectral lines from the secondary suggests it may be a white dwarf.

The orbital motion of a star in a gravitationally bound system can be detected in long-term astrometric measurements, which can also affect parallax determination. In the Gaia data, the quality of the astrometric solution is described by the renormalized unit weight error \citep[$RUWE$;][]{2018A&A...616A...2L}, with a detailed description in Section~14.1.2 of the Gaia DR2 documentation \citep{2018gdr2.reptE....V}. For a single star, $RUWE \sim 1.0$, whereas values greater than 1.4 may indicate perturbations, such as orbital motion in a multiple system.

The Gaia DR2 parallax of RY~Lep is $2.4632(503)$\,mas with $RUWE = 1.366$ \citep{2018A&A...616A...1G}. Using the Starhorse model and DR2 data, \citet{2019A&A...628A..94A} derived a distance of $d = 396^{+10}_{-10}$\,pc. The Gaia DR3 parallax is $2.2932(954)$\,mas \citep{2023A&A...674A...1G}, and the distance derived from Starhorse2 and EDR3 data is $d = 461^{+15}_{-86}$\,pc \citep{2022A&A...658A..91A}. The high $RUWE = 5.649$ in DR3 may reflect the influence of orbital motion on the astrometric solution.

Combined Hipparcos and Gaia DR2 measurements show a proper motion anomaly of $|\delta \mu_{\rm H/G2}| = 23.497(23)$\,mas\,yr$^{-1}$ \citep{2019A&A...623A..72K}, corresponding to a tangential velocity of $v_{\rm T} = 8.4(3)$\,km\,s$^{-1}$. Assuming a primary mass of $M_1 = 2.4(1)$\,M$_{\sun}$ from \citet{2000A&AS..141..371G}, \citet{2019A&A...623A..72K} estimated the secondary mass (normalized to an orbital radius of 1\,AU) as $M_2 = 0.51^{+0.14}_{-0.06}$\,M$_{\sun}$. They also derived the primary radius as $R_1 = 5.78(29)$\,R$_{\sun}$ using the surface brightness–color relation of \citet{2004A&A...426..297K}.

\begin{figure}
    \centering
    \includegraphics[width=0.45\textwidth]{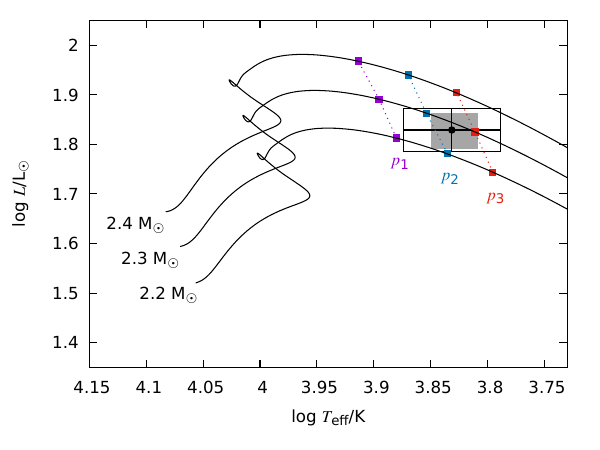}
    \caption{The HR diagram showing the position of RY~Lep. The open box encompasses the whole range of $T_{\rm eff}$ reported in the literature, while the gray box represents the values derived from our spectroscopic analysis. The evolutionary tracks were computed for $X_0 = 0.7$, 
    $Z = 0.008$ and $V_{\rm rot,0} = 45$\,km\,s$^{-1}$. 
    Lines of constant frequency $4.4415$\,d$^{-1}$ correspond to the radial fundamental mode ($p_1$), first overtone ($p_2$),
    and second overtone ($p_3$).}%
    \label{fig:rylep_hr_diagram}%
\end{figure}

In Figure~\ref{fig:rylep_hr_diagram}, we show the position of RY~Lep on the Hertzsprung–Russell (HR) diagram.
The determination of $(T_{\rm eff},~L)$ is described in Section \ref{sec:RY_Lep_photometry}. For reference, we also show evolutionary tracks for $M=2.2,~2.3,~2.4$\,M$_{\sun}$, computed with the Warsaw–New Jersey code \citep{1969AcA....19....1P, 1999AcA....49..119P} for three stellar masses, adopting OPAL opacities \citep{1996ApJ...464..943I}, solar chemical mixture from \citet{2009ARA&A..47..481A} (AGSS09), and the OPAL2005 equation of state \citep{1996ApJ...456..902R,2002ApJ...576.1064R}. Fig.\,\ref{fig:rylep_hr_diagram} also shows lines of constant frequency $4.4415$\,d$^{-1}$ corresponding to the first three radial modes ($p_1$, $p_2$, $p_3$). These lines intersect the error box of RY~Lep, with models indicating a hydrogen-shell burning (HSB) phase of evolution.
The other parameters were: the initial hydrogen abundance $X_0=0.70$, metallicity of $Z=0.008$, and the initial velocity of rotation $V_{\rm rot, 0} = 45$\,km\,s$^{-1}$. We assumed the mixing length parameter $\alpha_{\rm MLT} = 0.5$ and neglected overshooting from the convective core $\alpha_{\rm ov}=0.0$.


\section{Atmospheric parameters and chemical abundances from spectroscopy}\label{sec:RY_Lep_spectroscopy}

The available spectra of RY~Lep include one high-resolution SALT spectrum and a set of 852 low-resolution spectra obtained by \citet{2009MNRAS.394..995D} using the 2.3\,m telescope at Siding Spring Observatory (SSO). All spectra were normalized using the neural network–based Suppnet software \citep{2022A&A...659A.199R}. The radial velocity ($v_{\rm rad}$), effective temperature ($T_{\rm eff}$), overall metallicity ([m/H]), and chemical abundances were determined using the iSpec software \citep{2014A&A...569A.111B, 2019MNRAS.486.2075B}.

\subsection{SALT}
\begin{figure*}
	\centering
	\includegraphics[width=\textwidth]{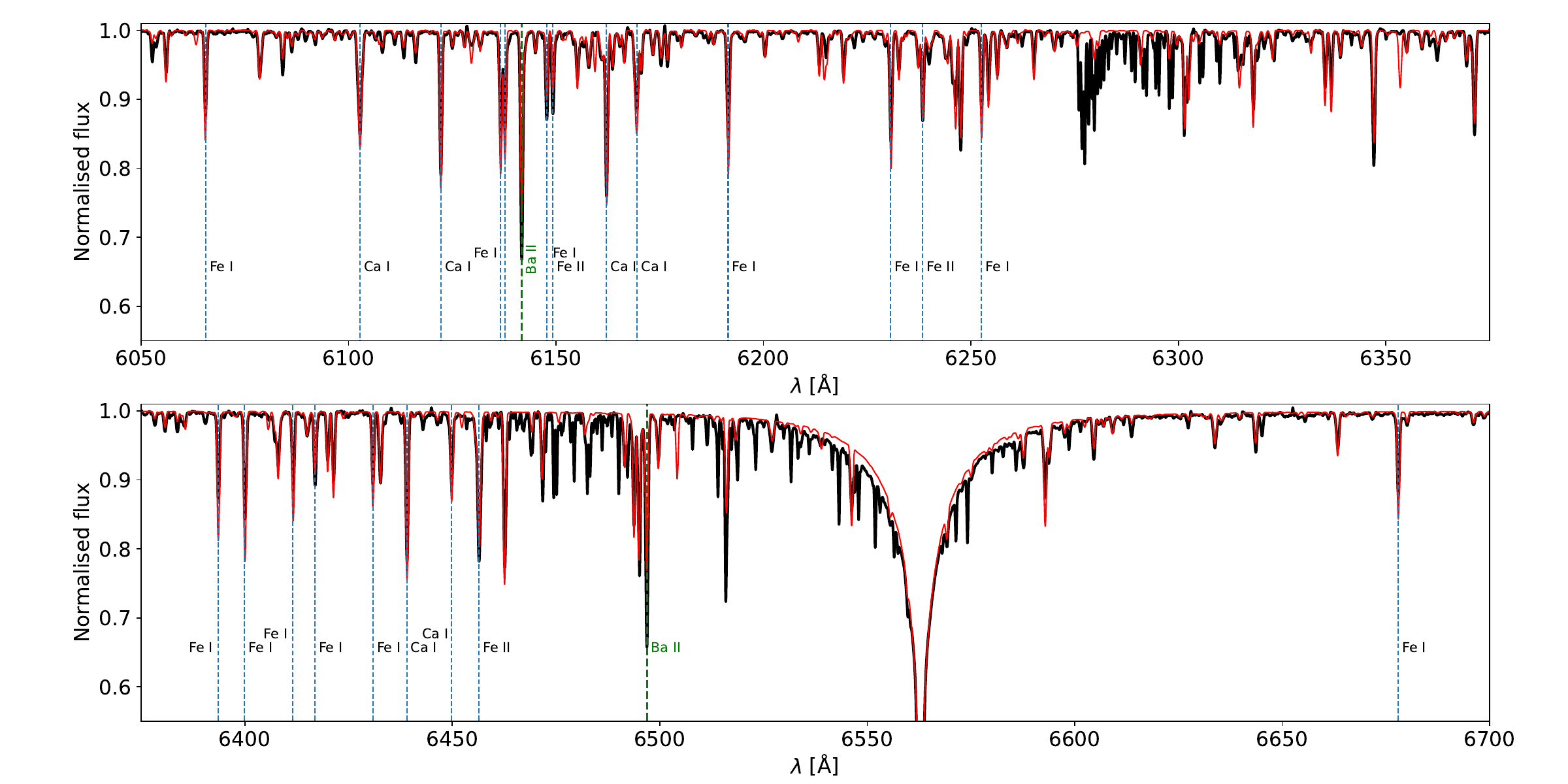}
	\caption{The normalized SALT spectrum of RY~Lep is shown as a black line. The synthetic spectrum with $T_{\rm eff}=6750$\,K,
		$\log{g}=3.5$, [m/H] = -0.4, $\xi=4$\,km\,s$^{-1}$, and $V_{\rm rot}\sin{i} = 20$\,km\,s$^{-1}$ is shown as 
		a red line. Vertical blue lines mark the wavelengths of Fe and Ca lines used to determine the overall metallicity, while the two green vertical lines indicate the positions of the Ba\,{\footnotesize II} spectral lines at 6141.7 and 6496.8 \AA.}%
	\label{fig:rylep_spectrum_SALT_part}%
\end{figure*}

The high-resolution SALT spectrum of RY~Lep \citep{2010SPIE.7737E..25C} was obtained over the wavelength range 3800–8000\,\AA\,with a resolving power of $R = 41\,987$. The observation was conducted on 9 February 2023 with an exposure time of 1200\,s,
corresponding to approximately 6\% of RY~Lep’s pulsation cycle. In Fig.~\ref{fig:rylep_spectrum_SALT_part}, we show a portion 
of the SALT spectrum covering 6050–6700\,\AA, which overlaps with the wavelength range of the SSO spectra. To determine the stellar parameters of RY~Lep, we used the full spectral range 4200–6700\,\AA, which is shown in Appendix~\ref{sec:appendix_spectrum} (Fig.~\ref{fig:rylep_spectrum_SALT}).

\begin{deluxetable*}{|r|l|r|r|r|l|r|r|} 
	\centering
	\tabletypesize{\scriptsize}
	\tablewidth{0pt}
	\tablecaption{Abundances of chemical elements in RY~Lep estimated from the SALT spectrum. \label{tab:RY_Lep_SALT_elements_abundances}}
	\tablehead{
		\colhead{$Z$ } & \colhead{  Element } & \colhead{  [X/H] } & \colhead{  $\sigma_{\rm [X/H]}$ } & \colhead{  [X/Fe] } & \colhead{  origin }
		& \colhead{  $N_s/N_{\rm tot}$} & \colhead{ $N_s/N_{\rm tot}$}  \\[-5pt]
		\colhead{} & \colhead{  } & \colhead{  } & \colhead{  } & \colhead{  } & \colhead{  } & \colhead{  A99 [\%] } & \colhead{  B14 [\%]} }
	\startdata
	6   &   C  &       -0.36   &         0.14  &        0.08  & $3\alpha$ & - &-\\ 
	7   &   N  &       -0.33   &         0.72  &        0.11  & CNO & - & -\\ 
	8   &   O  &       -0.10   &         0.26  &        0.34  & $3\alpha$ & - &-\\ 
	11   &   Na &       -0.06   &         0.13  &        0.38  & C burning & - &-\\ 
	12   &   Mg  &       -0.40   &         0.04  &       0.04  & O,Ne burning & - &-\\ 
	14   &   Si  &       -0.17   &         0.06  &       0.27  & O burning & - &-\\ 
	16   &   S   &       -0.10   &         0.12  &       0.34  & O burning & - &-\\ 
	20   &   Ca  &       -0.39   &         0.05  &        0.05  & O,Si burning & - &-\\ 
	21   &   Sc  &       -0.11   &         0.04  &        0.33  & O,Ne burning & - &-\\ 
	22   &   Ti  &       -0.24   &         0.03  &        0.20  & Si burning & - &-\\ 
	23   &   V   &        0.05   &         0.10  &        0.49  & Si burning & - &-\\ 
	24   &   Cr  &       -0.27   &         0.03  &        0.17  & Si burning & - &-\\ 
	25   &   Mn  &       -0.47   &         0.07  &       -0.03  & Si burning & - &-\\ 
	26   &   Fe  &       -0.44   &         0.01  &        0.00  & Si burning & - &-\\ 
	27   &   Co  &       -0.17   &         0.22  &        0.27  & Si burning & - &-\\ 
	28   &   Ni  &       -0.28   &         0.04  &        0.16  & Si burning & - &-\\ 
	29   &   Cu  &       -0.12   &         0.15  &        0.32  & Si burning    & 1 & -\\ 
	30   &   Zn  &       -0.18   &         0.12  &        0.26  & Si burning    & 0.9 & -\\ 
	38   &   Sr  &        0.29   &         0.07  &        0.73  & $s$-process   & 85  & 69(6)\\ 
	39   &   Y   &        0.34   &         0.06  &        0.78  & $s$-process   & 92  & 72(8)\\ 
	40   &   Zr  &        0.32   &         0.07  &        0.76  & $s$-process   & 83  & 66(7)\\ 
	56   &   Ba  &        0.40   &         0.07  &        0.84  & $s$-process   & 81  & 85(7) \\ 
	57   &   La  &        0.32   &         0.06  &        0.76  & $s$-process   & 62  & 76(5) \\ 
	58   &   Ce  &        0.52   &         0.04  &        0.96  & $s$-process   & 77  & 84(6)\\ 
	59   &   Pr  &        0.73   &         0.09  &        1.17  & $s/r$-process & 49  & 50(4) \\ 
	60   &   Nd  &        0.56   &         0.06  &        1.00  & $s$-process   & 56  & 58(4) \\ 
	62   &   Sm  &        0.72   &         0.06  &        1.16  & $r$-process   & 29  & 31(2) \\ 
	63   &   Eu  &        1.01   &         0.09  &        1.45  & $r$-process   & 5.8 & 6.0(4)\\ 
	64   &   Gd  &        0.74   &         0.14  &        1.18  & $r$-process   & 15  & 15(1) \\ 
	66   &   Dy  &        1.12   &         0.07  &        1.55  & $r$-process   & 15  & 15(1) \\ 
	\enddata
	\tablenotetext{}{The sixth column lists the dominant nucleosynthesis process, following \citet{2004AA...416.1117C} for elements with atomic number $Z \le 30$ and \citet{2014ApJ...787...10B} for elements with $Z > 30$. The last two columns give the contributions of the $s$-process to the total abundance of each element, as derived by \citet{1999ApJ...525..886A} (A99) and \citet{2014ApJ...787...10B} (B14), respectively.}
\end{deluxetable*}

The radial velocity of RY~Lep measured from the SALT spectrum is $v_{\rm rad} = 11.8(2)$\,km\,s$^{-1}$. The effective temperature, determined from the wings of the H$\alpha$, H$\beta$, and H$\gamma$ lines, is $T_{\rm eff} = 6878(15)$\,K. Using the equivalent widths of Fe\,{\footnotesize I, II} lines with well-determined oscillator strengths ($gf$) in the range 4620–4635\,\AA\ \citep{1998A&A...338.1041L}, we derived a surface gravity of $\log g = 3.51(5)$ and a microturbulent velocity of $\xi = 3.66(5)$\,km\,s$^{-1}$.

From well-isolated Mg, Ca, Ti, and Fe lines, we estimated a projected rotational velocity of $V_{\rm rot}\sin i = 21.9(3)$\,km\,s$^{-1}$, an overall metallicity of [m/H] = –0.40(1), and an $\alpha$-element abundance ratio of [$\alpha$/Fe] = 0.10(2). The macroturbulent velocity, $\zeta$, is degenerate with rotation. Since the empirical relation for $\zeta$ implemented in iSpec is valid only for $T_{\rm eff} < 6600$\,K, we adopted $\zeta = 0$. Spectral lines can also be broadened by pulsations, particularly for modes with periods shorter than the exposure time ($P \lesssim 1200$\,s), but at the resolving power of $R \approx 40{\,}000$, this effect is also largely degenerate with rotation.

\begin{figure}[ht]
	\centering
	\includegraphics[width=0.45\textwidth]{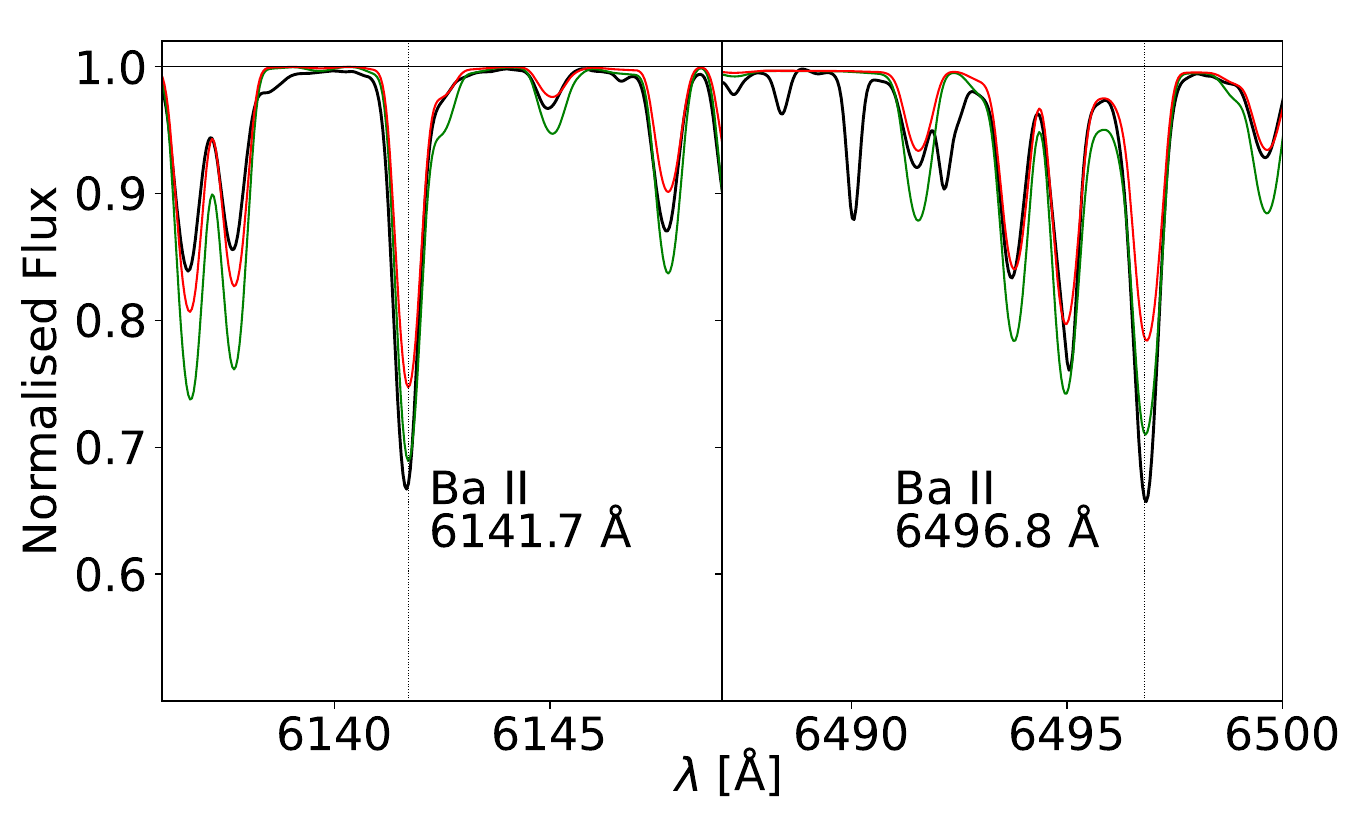}
	\caption{Barium (Ba\,{\footnotesize II}) lines in the SALT spectrum of RY~Lep. The black line shows the observed spectrum.
		Synthetic spectra were computed with a microturbulent velocity $\xi=4$\,km\,s$^{-1}$, projected rotational velocity $V_{\rm rot}\sin{i} = 21.9$\,km\,s$^{-1}$, and metallicities [m/H]=0.0 (green) and -0.4 (red).}%
	\label{fig:rylep_barium_spectrum_SALT}%
\end{figure}

\begin{figure*}[ht]
    \centering
    \includegraphics[width=\textwidth]{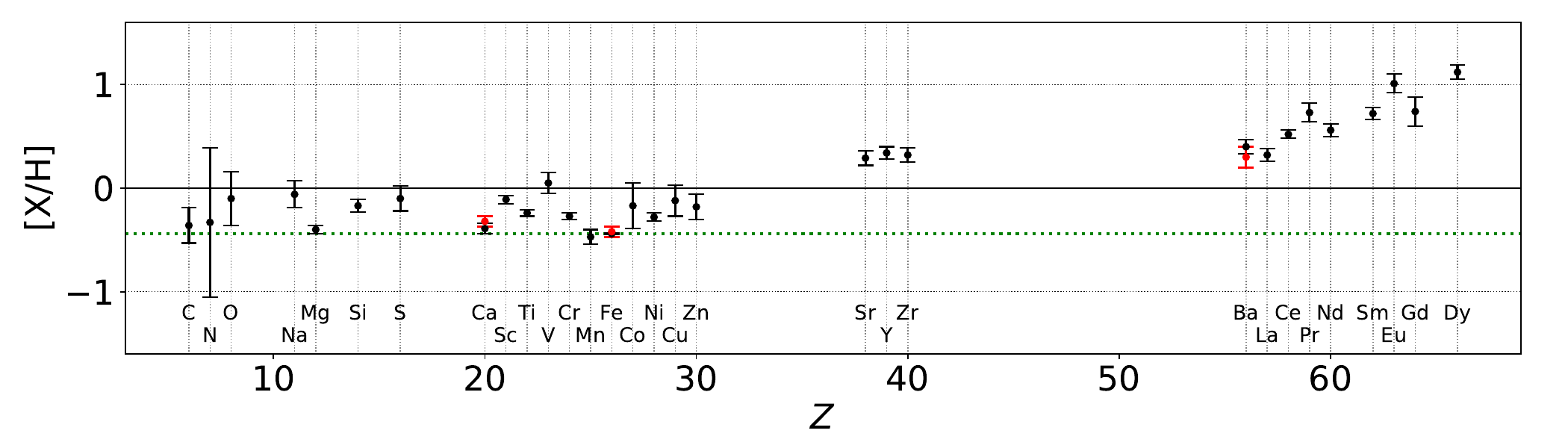}
    \caption{Abundances of chemical elements of RY~Lep derived from the SALT (black points) and SSO (red points) spectra.
    The horizontal green line marks the iron abundance, [Fe/H], from the SALT spectrum. }%
    \label{fig:rylep_SALT_abundances}%
\end{figure*}

We used the spectral lines from the SALT spectrum to determine individual chemical abundances. Two barium lines are highlighted in 
Fig.~\ref{fig:rylep_barium_spectrum_SALT}, and additional lines of overabundant elements are shown in Fig.~\ref{fig:rylep_spectrum_SALT_lines}. Lines affected 
by telluric contamination were excluded from the analysis. All measured elemental abundances are listed in Table~\ref{tab:RY_Lep_SALT_elements_abundances} and 
shown as black points in Fig.~\ref{fig:rylep_SALT_abundances} as a function of atomic number $Z$.


\subsection{SSO}
Using the atmospheric parameters $(\log g, \xi, V_{\rm rot}\sin i)$ derived from the SALT spectrum, we re-analyzed a set of 852 low-resolution spectra of RY\,Lep obtained by \citet{2009MNRAS.394..995D} with the SSO telescope.

\begin{figure*}
    \centering
    \includegraphics[width=\textwidth]{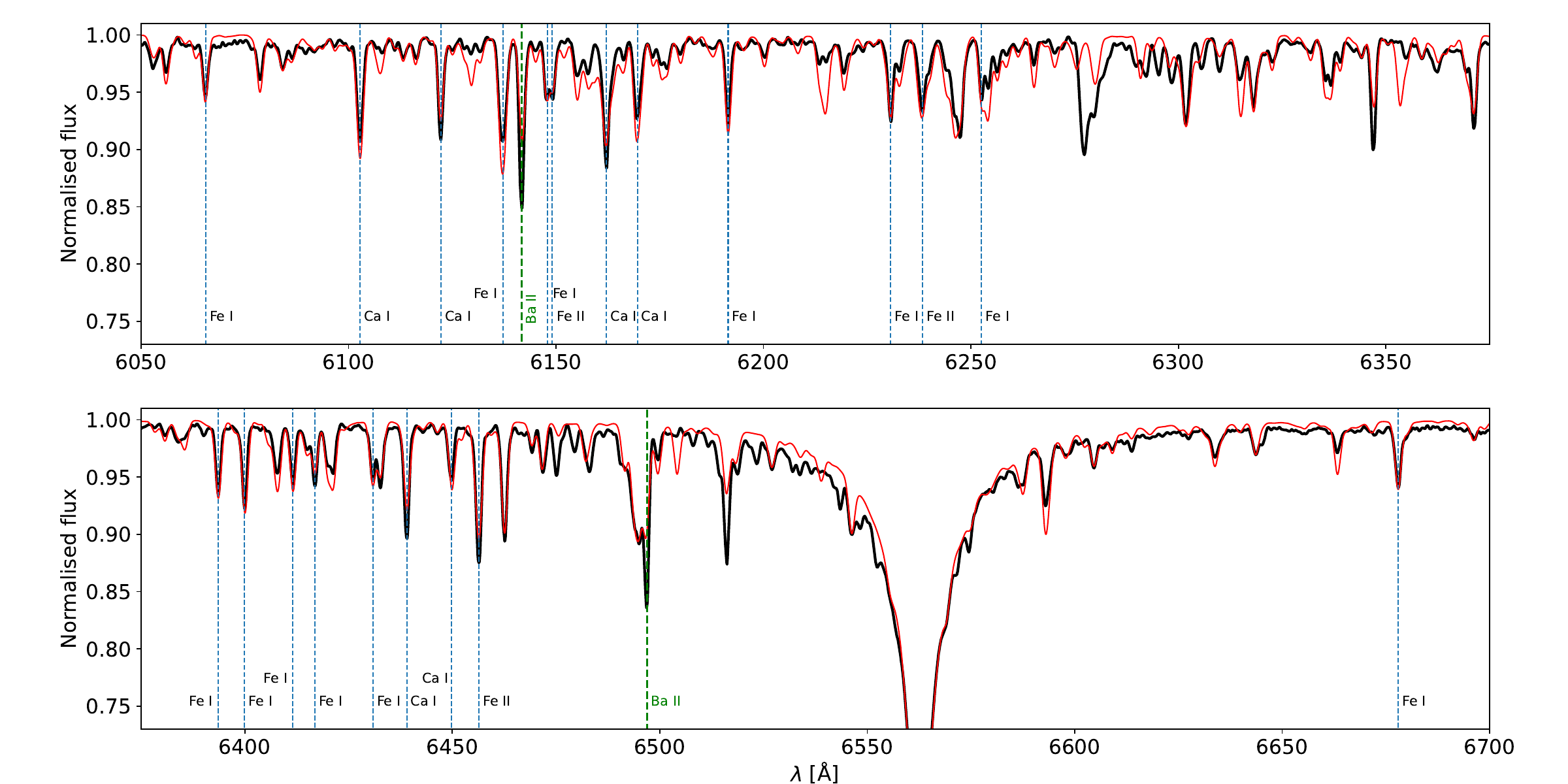}
    \caption{The median SSO spectrum of RY\,Lep for December 2005 is shown as a black line. The synthetic spectrum, computed with $T_{\rm eff} = 6750$\,K, $\log g = 3.5$, [m/H] = 0.0, $\xi = 4$\,km\,s$^{-1}$, and $V_{\rm rot}\sin i = 20$\,km\,s$^{-1}$, is plotted as a red line. Vertical blue lines indicate the wavelengths of Fe and Ca lines used to determine the overall metallicity. Two green vertical lines mark the positions of Ba\,{\footnotesize II} lines at 6141.7 and 6496.8\,\AA.}%
    \label{fig:rylep_spectra_2005}%
\end{figure*}

The observations were carried out with the red arm of the Double Beam Spectrograph using the 1200\,mm$^{-1}$ grating. RY\,Lep was observed over the wavelength range 5795--7010\,\AA, covering the H$\alpha$ line as well as strong lines of Ca, Fe, and Ba. The data were collected over six nights: two in February 2004, two in October 2004, and two in December 2005. Although the spectra were obtained in slightly different spectral settings, all include a common wavelength range of 6050--6700\,\AA. Figure~\ref{fig:rylep_spectra_2005} shows an example of the median spectrum constructed from all exposures taken in December 2005.

The wings of H$\alpha$ were used to determine the effective temperature. A typical uncertainty in $T_{\rm eff}$ is approximately 150\,K. The mean effective temperature is 6742\,K, with a range of 6430--7080\,K.

\begin{figure}
	\centering
	\includegraphics[width=0.45\textwidth]{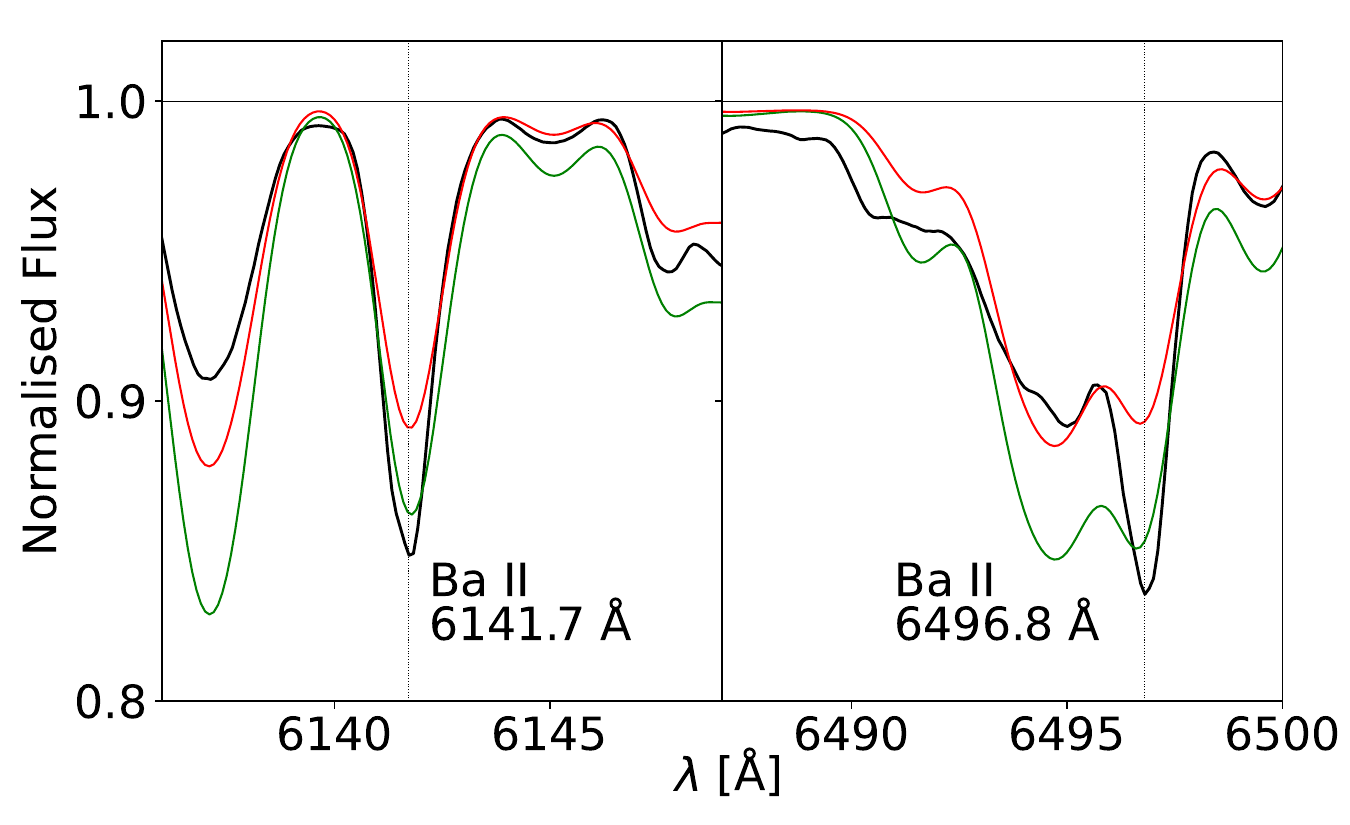}
	\caption{Barium (Ba\,{\footnotesize II}) lines in the SSO spectrum of RY\,Lep. The black line shows the median spectrum for December 2005. Synthetic spectra were computed with a microturbulent velocity $\xi=4$\,km\,s$^{-1}$, projected rotational velocity $V_{\rm rot}\sin{i} = 21.9$\,km\,s$^{-1}$, and metallicities [m/H]=0.0 (green) and -0.4 (red)}%
	\label{fig:rylep_barium_spectrum}%
\end{figure}

Next, we determined the overall atmospheric metallicity and the abundances of Ca, Fe, and Ba, which have well-separated spectral lines. To this end, a median spectrum was computed for each night. The adopted atmospheric parameters are $T_{\rm eff} = 6750$\,K, $\log g = 3.5$, $\xi = 3.66$\,km\,s$^{-1}$, and $V_{\rm rot}\sin i = 21.9$\,km\,s$^{-1}$. We obtained [Fe/H] = $-0.42(5)$, [Ca/H] = $-0.32(5)$, and [Ba/H] = $0.3(1)$. These abundances are shown as red points in Fig.~\ref{fig:rylep_SALT_abundances} and are consistent, within the uncertainties, with the values derived from the SALT spectrum.
In all analyzed spectra, two strong lines of Ba\,{\footnotesize II} at 6141.7\,\AA\ and 6496.9\,\AA\ were detected. Both lines in the median spectrum for December 2005 are presented in Fig.~\ref{fig:rylep_barium_spectrum}.

\section{Atmospheric parameters from photometry}\label{sec:RY_Lep_photometry}

The atmospheric parameters of RY~Lep have been determined multiple times from multicolor photometry. 
\citet{1985A&A...149..465D} derived an effective temperature in the range $T_{\rm eff} = 6150$–$7480$\,K
and a surface gravity $\log g = 3.27$–3.55 from Walraven VBLUW photometry, estimating a metallicity [Fe/H] $= +0.1(1)$ 
using the method of \citet{1981A&A....99L...1P}. \citet{1990Ap&SS.169..113R} obtained $T_{\rm eff} = 7090$\,K and 
$\log g = 3.38$ from Strömgren photometry using the \citet{1972A&A....17..367P} calibration, with metallicity [m/H] $= +0.2$.
Using the same observations, \citet{1995MNRAS.277..965R} found $T_{\rm eff}$ varying between 6680 and 7540\,K 
and $\log g$ between 3.31 and 3.58, with [m/H] $= +0.27$. \citet{1997PASP..109.1221M} derived mean values 
of $T_{\rm eff} = 7115$\,K and $\log g = 3.42$ using Kurucz model atmospheres. \citet{2009A&A...501..941H}, 
based on Strömgren photometry \citep{1991PASP..103..494P} and the calibration of \citet{2007A&A...475..519H},
estimated $T_{\rm eff} = 6980$\,K and [Fe/H] $= -0.03$. \citet{2006ApJ...638.1004A} derived $T_{\rm eff} = 6870^{+54}_{-637}$\,K 
from Hipparcos/Tycho-2 $BV$ photometry combined with 2MASS $JHK$ data, while \citet{2016A&A...591A.118S} estimated 
$T{\rm eff} = 6849(105)$\,K using angular diameters and total fluxes. \citet{2012MNRAS.427..343M} found $T_{\rm eff} = 6788$\,K 
from spectral energy distribution analysis.

We used Strömgren photometry from \citet{1989PhDT.......139R}, collected in a single night in 1988 with 
the 50-cm Danish telescope at ESO. From Kurucz models, we estimated $T_{\rm eff} = 6590$–7160\,K, $\log g = 3.12$–3.41,
and [m/H] = –0.07(4). Using the Strömgren photometry of \citet{2004CoAst.145...48R}, 
we derived $T_{\rm eff} = 6835$–7345\,K, $\log g = 3.35$–3.57, and [m/H] $= -0.22(12)$. From the $uvby$ data 
of \citet{1991PASP..103..494P}, we obtained $T_{\rm eff} = 6855(25)$\,K, $\log g = 3.25(4)$, and [m/H]$= -0.05(4)$.
The Strömgren $\beta$ index in these data also allows determination of [Fe/H] using the relation of \citet{1990A&A...236..440B}
and standard calibrations for AF-type stars \citep{1975AJ.....80..955C, 1979AJ.....84.1858C}. We obtained
[Fe/H]$= 0.06(9)$ and $0.05(3)$ from \citet{1989PhDT.......139R} and \citet{1991PASP..103..494P}, respectively.
Using the [m/H]-$\delta m_1$ relation for A3-F0 stars ($2.72 < \beta < 2.88$) \citep{1993A&A...274..391S}, 
we determined [m/H]$=-0.03(6)$ and $0.05(3)$ for the 1988 normal points and the \citet{1991PASP..103..494P} data, respectively.

\begin{deluxetable*}{llllllll} 
\centering
    \tabletypesize{\scriptsize}
\tablewidth{0pt}
			\tablecaption{Stellar parameters of RY~Lep derived in this work using photometric and spectroscopic data.\label{tab:RY_Lep_teff_logg_mh_feh_table}}
				\tablehead{
\colhead{~~~~~~~Data} & \colhead{$T_{\rm eff}$} & \colhead{$\log g$} & \colhead{[m/H]} & \colhead{[Fe/H]}  & \colhead{$\xi$}
& \colhead{$V_{\rm rot}\sin{i}$}
\\[-9pt]
\colhead{} & \colhead{   [K]~     } & \colhead{  [cgs]} & \colhead{            } & \colhead{             } & \colhead{[km\,s$^{-1}$]}
& \colhead{}
} 
				\startdata
				SALT & 6878(15) & 3.51(5) & $-0.40(1)$ & $-0.44(1)$  &3.66(5) & 21.9(3)\\
				SSO, Derekas 2004-2005 & [6430,7080] & -- & $-0.3(1)$ & $-0.35(10)$  & -- & -- \\
				\hline
				$uvby$ Rodr\'iguez 1988 & [6590,7160] & [3.12, 3.41] & $-0.07(4)$& 0.06(9) & -- & -- \\
				$uvby$ Perry 1991 & 6855(25) & 3.25(4) & 0.01(5) & 0.05(3)  & -- & -- \\
				$uvby$ Rodr\'iguez 1998-2002 & [6835,7345] & [3.35, 3.57] & $-0.22(12)$ & --& --& -- \\               
		                \hline
				$\beta$ Rodr\'iguez 1988 & -- & -- & $-0.03(6)$ & -- & -- & -- \\
				$\beta$ Perry 1991 & -- & -- & 0.05(3) & -- & -- & -- \\
				\hline
\enddata

\end{deluxetable*}
\begin{figure*}
	\centering
	\includegraphics[width=\textwidth]{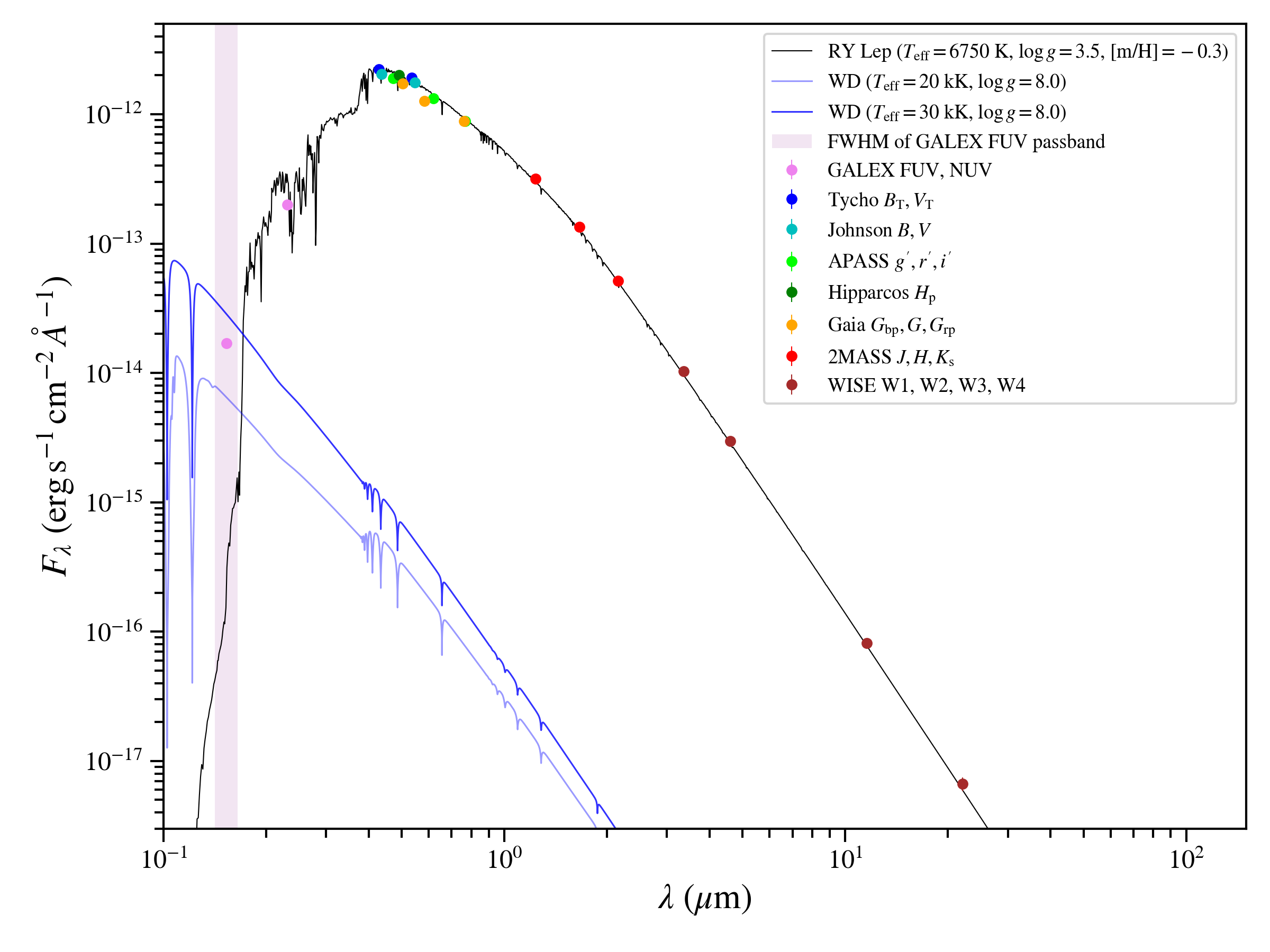}
	\caption{Spectral energy distribution of RY~Lep from various sources (see the legend). The black line shows a Kurucz atmospheric model for the star with $R = 5.7$\,R$_\odot$. The two blue lines represent spectra of a white dwarf with $R = 0.02$\,R$_\odot$.}%
	\label{fig:rylep_sed}%
\end{figure*}

To derive the luminosity, we adopted the full range $T_{\rm eff} = 6150$–7540\,K and $\log g = 3.12$–3.58 to interpolate
bolometric corrections (BC) from Kurucz models. 
Although the SALT spectra indicate a microturbulent velocity of $\xi \approx 4$\,km\,s$^{-1}$, Kurucz models are available for this velocity only at solar metallicity. For non-solar metallicities, we therefore adopted the grids computed at $\xi = 2$\,km\,s$^{-1}$, considering 
[m/H] = –0.5, –0.3, and –0.2.

Extreme BC values were used to estimate the luminosity uncertainty. Interstellar extinction $A_V = 0.0927$ was taken
from the {\it Bayestar2019} map \citep{2019ApJ...887...93G}.
Due to the high Gaia DR3 renormalized unit-weight error ($RUWE = 5.649$), we adopted a distance $d = 396(10)$\,pc from
the Starhorse model \citep{2019A&A...628A..94A}. This yields $\log (L/{\rm L}_\odot) = 1.829(43)$.
Using the same procedure, but adopting only the range $T_{\rm eff} = 6430$–7080\,K and $\log g = 3.46$–3.56 derived from spectroscopy, we obtained $\log(L/{\rm L_{\sun}}) = 1.827(37)$.

All atmospheric parameters derived in this work, both from spectroscopy and photometry, are listed 
in Table~\ref{tab:RY_Lep_teff_logg_mh_feh_table}.

Figure~\ref{fig:rylep_sed} shows the spectral energy distribution of RY~Lep, using photometry from GALEX, Tycho, Johnson,
Hipparcos, APASS, Gaia, 2MASS, and WISE. A Kurucz model for the primary and two synthetic white dwarf spectra
\citep{2010MmSAI..81..921K, 2009ApJ...696.1755T} are overplotted. A model with $R = 5.7$\,R$_\odot$ and $T_{\rm eff} = 6750$\,K 
reproduces the observed magnitudes across most wavelengths, except in the GALEX far-UV band, where a significant flux excess
is detected. This excess can be explained by a white dwarf with $T_{\rm eff} \approx 30$\,kK.

\section{The Fourier analysis}\label{sec:Fourier_analysis}
RY\,Lep has been the subject of numerous time-series observations, both ground-based and from space. In this section, we analyze the TESS, ASAS, and SuperWASP data sets, which have not been studied previously. 
We performed a Fourier analysis of the light curves and applied a standard pre-whitening procedure. The following subsections provide a detailed discussion, and Appendix~\ref{sec:appendix_frequencies} lists all detected frequencies.

\subsection{TESS}

High-quality photometric observations of RY\,Lep are available from the TESS satellite \citep{2015JATIS...1a4003R}. The star was observed in three sectors: S06 (TBJD 1468--1490), S32 (TBJD 2174--2200), and S33 (TBJD 2201--2227). Sector S06 is separated from sectors S32 and S33 by approximately two years and was therefore analyzed separately. The data from S32 and S33 were combined into a single time series to improve frequency resolution. For the analysis, we used 120-second cadence observations of the PDCSAP flux. The Rayleigh resolution was $\Delta\nu_R = 1/T = 0.046$\,d$^{-1}$ for S06 and 0.019\,d$^{-1}$ for the combined S32+S33 data.
The TESS light curves were first normalized by dividing each by its median value, and data points with a quality flag different from zero were excluded. 

Frequencies with a signal-to-noise ratio (S/N) greater than 5.0 were accepted. This threshold follows the estimates of \citet{2021AcA....71..113B} and corresponds to a false alarm probability (FAP) of 0.1\% for 120-second cadence TESS data. 
The noise level was estimated as the mean amplitude within a 1\,d$^{-1}$ window centered on each detected frequency.
This quantity should not be interpreted as the pure photometric noise; rather, it reflects an effective local background that includes both
the intrinsic photometric noise of the TESS data and residual stellar variability remaining after pre-whitening.
In particular, the elevated background level may arise from long-term phase modulation caused by the light-time effect in the binary system, as well as from intrinsic amplitude and/or frequency variability of the pulsation modes. These effects can leave unresolved residual power around the main frequencies, thereby increasing the measured local noise level.

\begin{figure}[ht]
	\centering
	\includegraphics[width=0.45\textwidth]{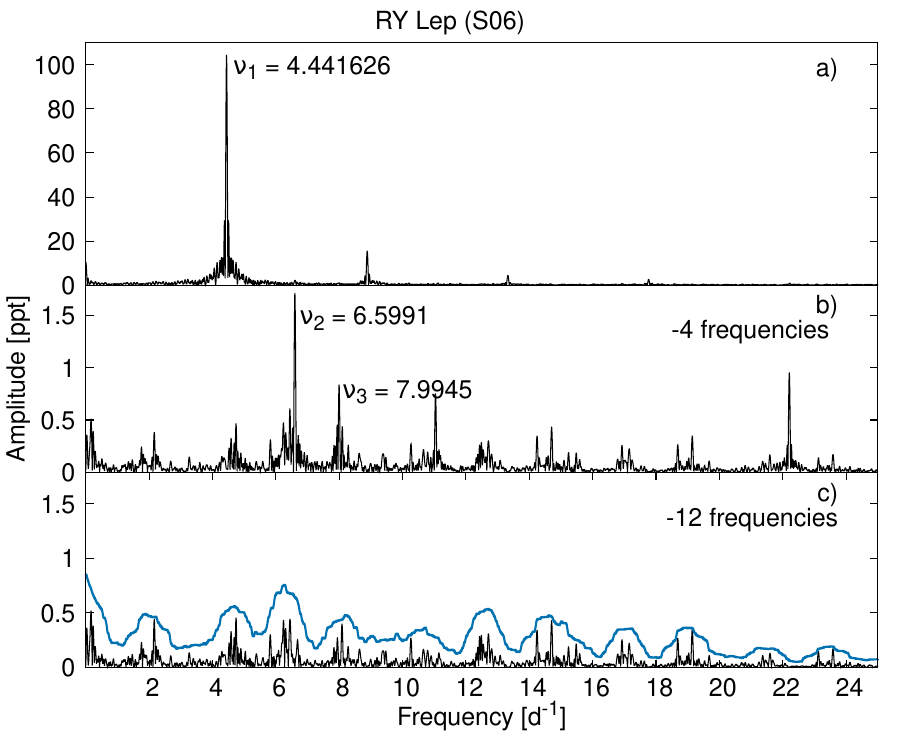}
	\caption{ Fourier amplitude periodograms derived from the TESS light curve of RY~Lep from Sector S06. From top to bottom, the panels show the periodogram of original data, after subtraction of four frequencies, and after subtraction of twelve frequencies. The blue horizontal line marks the signal-to-noise (S/N) threshold of 5.}%
	\label{fig:periodogram_TESS_S06}%
\end{figure}

In the S06 data, we identified 12 significant frequency peaks: three independent frequencies, $\nu_1=4.441626(4)$\,d$^{-1}$, $\nu_2=6.5991(3)$\,d$^{-1}$, and $\nu_3=7.9945(6)$\,d$^{-1}$; one combination frequency at 11.0399\,d$^{-1}$; and eight harmonics of the dominant frequency. The frequency $\nu_2 \approx 6.6$\,d$^{-1}$ was previously reported by \citet{2004CoAst.145...48R} and \citet{2009MNRAS.394..995D}, while $\nu_3$ is detected here for the first time. All detected frequencies, along with their amplitudes, signal-to-noise ratios, and identifications, are listed in Table~\ref{freq_RY_Lep_S06}. The Fourier amplitude periodograms for the original data and after two steps of pre-whitening are shown in Fig.~\ref{fig:periodogram_TESS_S06}.

\begin{figure}
	\centering
	\includegraphics[width=0.45\textwidth]{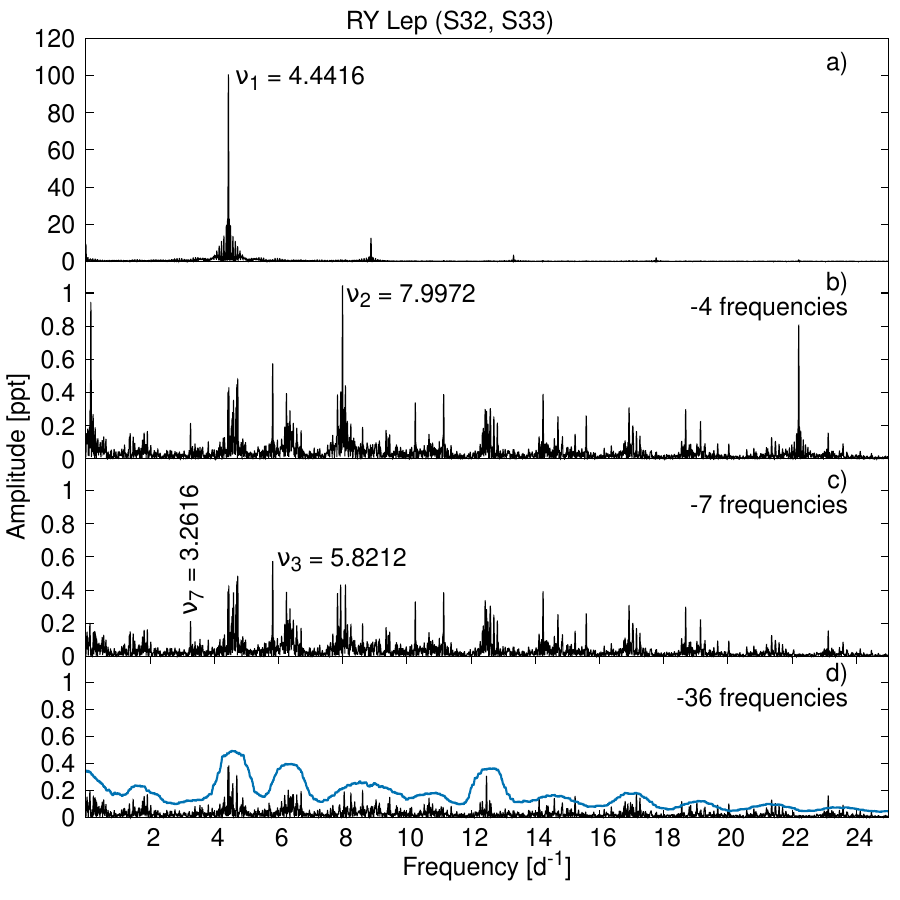}
	\caption{Fourier amplitude periodograms derived from the TESS light curve  of RY~Lep  from the combined S32 and S33 sectors. From top to bottom, the panels show the periodogram for the original data, after subtracting four terms, after subtracting 7 terms and after subtracting 36 terms. The blue horizontal line marks the signal-to-noise (S/N) threshold of 5.}%
	\label{fig:periodogram_TESS_S32_S33}%
\end{figure}

In the combined S32+S33 data, we identified 36 significant frequencies, including seven independent frequencies: $\nu_1 = 4.441552(1)$\,d$^{-1}$, $\nu_2 = 7.9972(1)$\,d$^{-1}$, $\nu_3 = 5.8212(2)$\,d$^{-1}$, $\nu_4 = 7.9394(2)$\,d$^{-1}$, $\nu_5 = 8.0866(2)$\,d$^{-1}$, $\nu_6 = 7.8392(3)$\,d$^{-1}$, and $\nu_7 = 3.2616(5)$\,d$^{-1}$. We also detected a low-frequency peak at $\nu_{\rm low}$ = 0.1571(1)\,d$^{-1}$, corresponding to a period of 6.36\,d. In addition, we identified eight harmonics of $\nu_1$ and one harmonic of $\nu_2$. Ten frequencies correspond to linear combinations of the independent peaks, while a further nine lie very close to the expected combination values, deviating by $1/T$ to $2.5/T$, and were therefore classified as combination frequencies. The Fourier amplitude periodograms for the original combined S32+S33 data and after three steps of pre-whitening are shown in Fig.~\ref{fig:periodogram_TESS_S32_S33}.
In Figures \ref{fig:periodogram_TESS_S06} and \ref{fig:periodogram_TESS_S32_S33}, residual signals are visible on both sides of some pre-whitened central peak. These features may arise from the effects mentioned above.

Notably, the 6.5991\,d$^{-1}$ frequency, detected in S06 and other datasets, was not present in the combined S32+S33 data. The frequency $\nu_2 = 7.9972$\,d$^{-1}$ corresponds to $\nu_3$ identified in Sector S06. Additionally, three independent frequencies were detected near $\nu_2$, each separated by more than $2.5/T$, which may indicate amplitude or frequency modulation of $\nu_2$ between the two sectors. Consequently, we also analyzed sectors S32 and S33 separately. The frequencies $\nu_1 = 4.441552(1)$\,d$^{-1}$, $\nu_2 = 7.9972(1)$\,d$^{-1}$, $\nu_3 = 5.8212(2)$\,d$^{-1}$, and low-frequency peak of about 0.16\,d$^{-1}$ were confirmed in each individual sector. All frequencies found in the combined S32+S33 data are presented in Table~\ref{freq_RY_Lep_S32_S33}, while the frequencies detected separately in S32 and S33 are listed in Tables~\ref{freq_RY_Lep_S32} and~\ref{freq_RY_Lep_S33}, respectively. The Fourier amplitude periodograms for S32 and S33 are shown in Figs.~\ref{fig:periodogram_TESS_S32} and~\ref{fig:periodogram_TESS_S33}, respectively.

The dominant frequency at 4.4416\,d$^{-1}$ was consistently detected across all datasets. A significant decrease in this frequency was observed over time, with a difference of $6 \times 10^{-5}$\,d$^{-1}$ between Sectors S06 and S32, and $1.5 \times 10^{-5}$\,d$^{-1}$ between Sectors S32 and S33. This change is likely associated with orbital motion within a binary system and corresponds to radial velocity differences of approximately 4\,km\,s$^{-1}$ between S06 and S32, and 1\,km\,s$^{-1}$ between S32 and S33.

A frequency near 7.995\,d$^{-1}$ was detected in all TESS sectors, although both its amplitude and value varied. Its value increased by approximately 0.003\,d$^{-1}$ between Sectors S32 and S33, and by 0.002\,d$^{-1}$ between Sectors S06 and S32. The frequency $\nu_5 = 8.087$\,d$^{-1}$ was detected only in the combined S32+S33 data and in the third segment of the SuperWASP observations (see Sect.\,5.3).

The low-frequency signal 0.16\,d$^{-1}$ is also visible in the periodogram of sector S06. However,  the signal-to-noise ratio of about 3.5 is below the adopted threshold of S/N = 5. \citet{2013MNRAS.431.2240B} also detected low-frequency peaks in the Kepler light curves of about 40\% of A-type stars. These peaks may be associated with spots or chemical inhomogenities that modulate the stellar light through rotation. If 0.16\,d$^{-1}$ corresponds to the surface rotation frequency, then adopting a radius of 5.7\,R$_{\sun}$ \citep{2019A&A...623A..72K} yields a rotational velocity of about 45\,km\,s$^{-1}$. Using $V_{\rm rot}\sin{i} = 21.9$\,km\,s$^{-1}$ derived from spectroscopy, we estimate an inclination angle of $30^{\circ}$.

The four frequencies detected in combined S32 and S33 data ($\nu_2 = 7.9972(1)$\,d$^{-1}$, $\nu_4 = 7.9394(2)$\,d$^{-1}$, $\nu_5 = 8.0866(2)$\,d$^{-1}$, and $\nu_6 = 7.8392(3)$\,d$^{-1}$) resemble components of a multiplet. However, since they are not uniformly spaced, they cannot be interpreted as belonging to a single rotationally split multiplet.  Among these frequencies, two pairs can be distinguished with similar separations: $\nu_2-\nu_6 = 0.158$\,d$^{-1}$ and $\nu_5-\nu_4 = 0.147$\,d$^{-1}$. 

As shown later, all seismic models of RY~Lep are in the post-main sequence phase of evolution. In such models, all nonradial modes, even at frequencies of $\sim 8$\,d$^{-1}$, are predominantly gravity modes \citep{2023MNRAS.526.1951D}.
 The Ledoux constant is then $C_{n\ell} \in (0.3,0.5)$ for $\ell = 1$ 
and $C_{n\ell} \approx 0.16$ for $\ell = 2$.
Using the first-order formulae  $\nu_{\rm split}=\nu_{\rm rot}(1-C_{n\ell})$, we get $\nu_{\rm rot} \in (0.23, 0.32)$ for $\ell = 1$ modes and $\nu_{\rm rot} \approx 0.19$ for $\ell = 2$ modes. These values of $\nu_{\rm rot}$ would correspond to the deeper layers probed by the gravity modes.  Thus, if we do observe the rotationally split modes and $\nu_{\rm low}$ corresponds to the surface rotation frequency, one can conclude that the star exhibits differential rotation.

\subsection{ASAS}
RY\,Lep was also observed as part of the All Sky Automated Survey (ASAS) between November 2000 and December 2009. These observations, obtained in the Johnson $V$ passband, provided a Rayleigh resolution of 0.0003\,d$^{-1}$. Among the available apertures, aperture 3 yielded the highest signal-to-noise ratio and was therefore selected for detailed frequency analysis. We used 613 points flagged as best or mean data. Because the temporal baseline of the ASAS data is substantially longer than that of the TESS sectors, the light-time effect significantly affects the photometric solution.

From the full dataset, we identified two independent frequencies: 4.441543(3)\,d$^{-1}$ and 6.59953(3)\,d$^{-1}$. Additionally, a frequency at 6.4450(3)\,d$^{-1}$ was detected, which is most likely an 2\,d$^{-1}$ alias of the dominant mode.

Because of the long gaps in the ASAS time series, we performed a supplementary analysis by dividing the dataset into five shorter segments, each spanning 396--680 days. In all segments, a frequency near 4.44\,d$^{-1}$ was consistently detected. Notably, in the second segment (July 2002–May 2004), the independent frequency at 6.5998(1)\,d$^{-1}$ was also present. The results of the Fourier analysis of the ASAS data are summarized in Table~\ref{freq_RY_Lep_ASAS}.

\subsection{SuperWASP}
We also utilized SuperWASP observations of RY\,Lep, spanning from October 2006 to March 2012. Although the total time baseline of 1963 days is substantial, the data are unevenly distributed. As in the case of the ASAS data, the light-time effect has a significant impact on the long-baseline SuperWASP photometric solution. Three distinct segments contain a large number of data points and are separated by approximately two years, with lengths of 162 days (October 2007--March 2008), 157 days (September 2009--March 2010), and 153 days (September 2011--March 2012). Outside these segments, the number of observations was insufficient to perform reliable Fourier analyses; consequently, we analyzed these three segments independently. The Rayleigh resolution within each densely sampled segment is approximately 0.006\,d$^{-1}$. The total number of all data points is 6201.

In the first segment, two independent frequencies, 4.44154(4)\,d$^{-1}$ and 6.5994(3)\,d$^{-1}$, were detected. Frequencies at 2.0015(2)\,d$^{-1}$, 5.0045(5)\,d$^{-1}$, and 10.0297(7)\,d$^{-1}$ were identified as daily aliases. In the second segment, only a single frequency of 4.4420(3)\,d$^{-1}$ exceeded the S/N threshold of 4.0. In the third segment, independent frequencies of 4.44161(4)\,d$^{-1}$, 6.6005(1)\,d$^{-1}$, and 8.087(1)\,d$^{-1}$ were detected. The dominant frequency remains consistent within the uncertainties across all segments. The amplitude of the 6.6\,d$^{-1}$ frequency more than doubled between the first and third segments, while its frequency decreased by 0.0011\,d$^{-1}$. All frequencies identified from the SuperWASP data are listed in Table~\ref{freq_RY_Lep_superwasp}.

\section{Mode identification}\label{sec:mode_identification}
The analysis of multiple observational datasets revealed light variations at several frequencies, which we associate with stellar pulsations. These frequencies, along with their corresponding amplitudes and phases, can be used for asteroseismic modeling. However, prior to such modeling, it is necessary to identify the pulsation modes.

To identify the mode degree $\ell$ of the dominant frequency, we applied the method of \citet{2003A&A...407..999D}, which determines $\ell$ simultaneously with two pulsational parameters by using the photometric amplitudes and phases measured in at least three passbands. The first parameter is the intrinsic mode amplitude, $\varepsilon$, multiplied by a factor that depends on the inclination angle $i$ and is expressed through the spherical harmonic $Y_\ell^m(i,0)$. Within linear theory, $\varepsilon$ defines the local radial displacement of a surface element produced by the pulsation.
The second parameter, the nonadiabatic parameter $f$, describes the ratio of the relative bolometric flux variation to the relative radial displacement of the stellar surface caused by the pulsation. Theoretical values of $f$ are obtained from linear nonadiabatic pulsation computations, whereas the intrinsic amplitude $\varepsilon$ remains undetermined within linear theory. Both $\varepsilon$ and $f$ are complex quantities, each having an amplitude and a phase.

With photometric amplitudes and phases measured in at least three passbands, the goodness of fit for a given spherical degree 
$\ell$ is quantified using the following discriminant:

\begin{equation}
	\chi^2 = \frac{1}{2N - N_p} \sum_{i=1}^{N} \frac{ \left| \mathcal{A}^{\rm obs}_{\lambda_i} - \mathcal{A}^{\rm cal}_{\lambda_i} \right|^2 }{ \sigma^2_{\lambda_i} },
\end{equation}
where $\mathcal{A}^{\rm obs}_{\lambda_i}$ and $\mathcal{A}^{\rm cal}_{\lambda_i}$ are the observed and calculated complex amplitudes in the $i$-th passband, respectively.  The uncertainties  $\sigma_{\lambda_i}$ 
in the observed complex amplitudes are:
\begin{equation}
	|\sigma_\lambda|^2= \sigma^2 (A_{\lambda})  +  A_{\lambda}^2 \sigma^2(\varphi_\lambda).
\end{equation}
The total number of observations is $2N$ (amplitudes plus phases for $N$ passbands), and the number 
of fitted parameters is $N_p = 4$, since both $\tilde{\varepsilon}=\varepsilon Y_\ell^m(i,0)$ and $f$ are complex.
If a clear minimum is found in the difference between observed and calculated amplitudes and phases, 
then the corresponding degree $\ell$, together with the associated complex values of $\tilde\varepsilon$ and $f$, 
can be taken as the most probable mode identification.

We used the amplitudes and phases in the Str\"omgren $uvby$ filters (see Table \ref{tab:RY_Lep_uvby_Rodriguez}) and applied the method of \citet{2003A&A...407..999D}. The flux derivatives with respect to $\log T_{\rm eff}$ and $\log g$, which appear in the expression for
$\mathcal{A}^{\rm cal}_{\lambda_i}$, were computed from Vienna  model atmospheres \citep{2002A&A...392..619H}, which include the turbulent-convection treatment of \citet{1996ApJ...473..550C}. For the limb darkening law, we computed coefficients assuming 
the non-linear formula of \citet{2000A&A...363.1081C}.

\begin{figure*}
	\centering
	\includegraphics[width=0.95\textwidth]{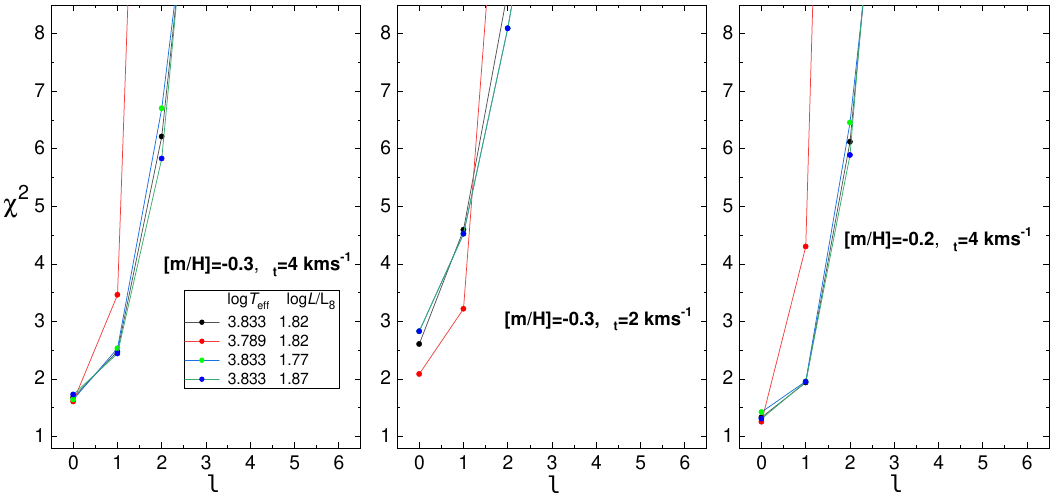}
	\caption{The discriminant $\chi^2$ as a function of $\ell$ for the dominant frequency 4.4415\,d$^{-1}$ of RY~Lep. There is shown the effect of effective temperature, luminosity, metallicity, and microturbulent velocity.}
	\label{fig:rylep_chi_ell}%
\end{figure*}

In Fig.~\ref{fig:rylep_chi_ell}, we plot the values of $\chi^{2}$ as a function of $\ell$ for the dominant frequency $4.4415\,\mathrm{d}^{-1}$ of RY~Lep. Five pairs of $(T_{\rm eff},\,L)$ from the edges of the error box were considered, and the effects of metallicity and microturbulent velocity were examined. As shown in the figure, the minimum value of $\chi^{2}$ indicates a radial mode in each case.

\begin{figure*}
	\centering
	\includegraphics[width=0.48\textwidth,height=7.1cm]{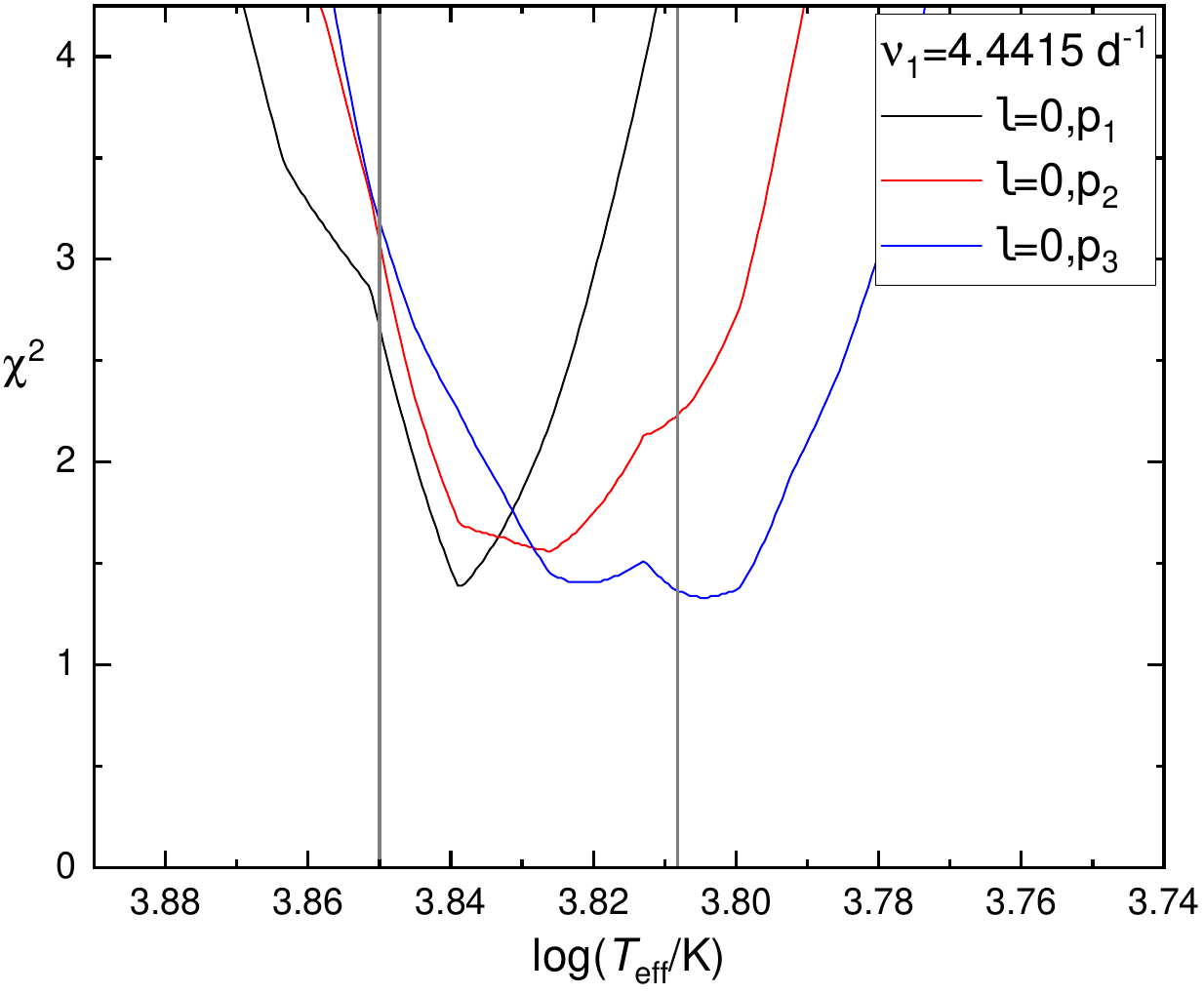}
	\includegraphics[width=0.48\textwidth]{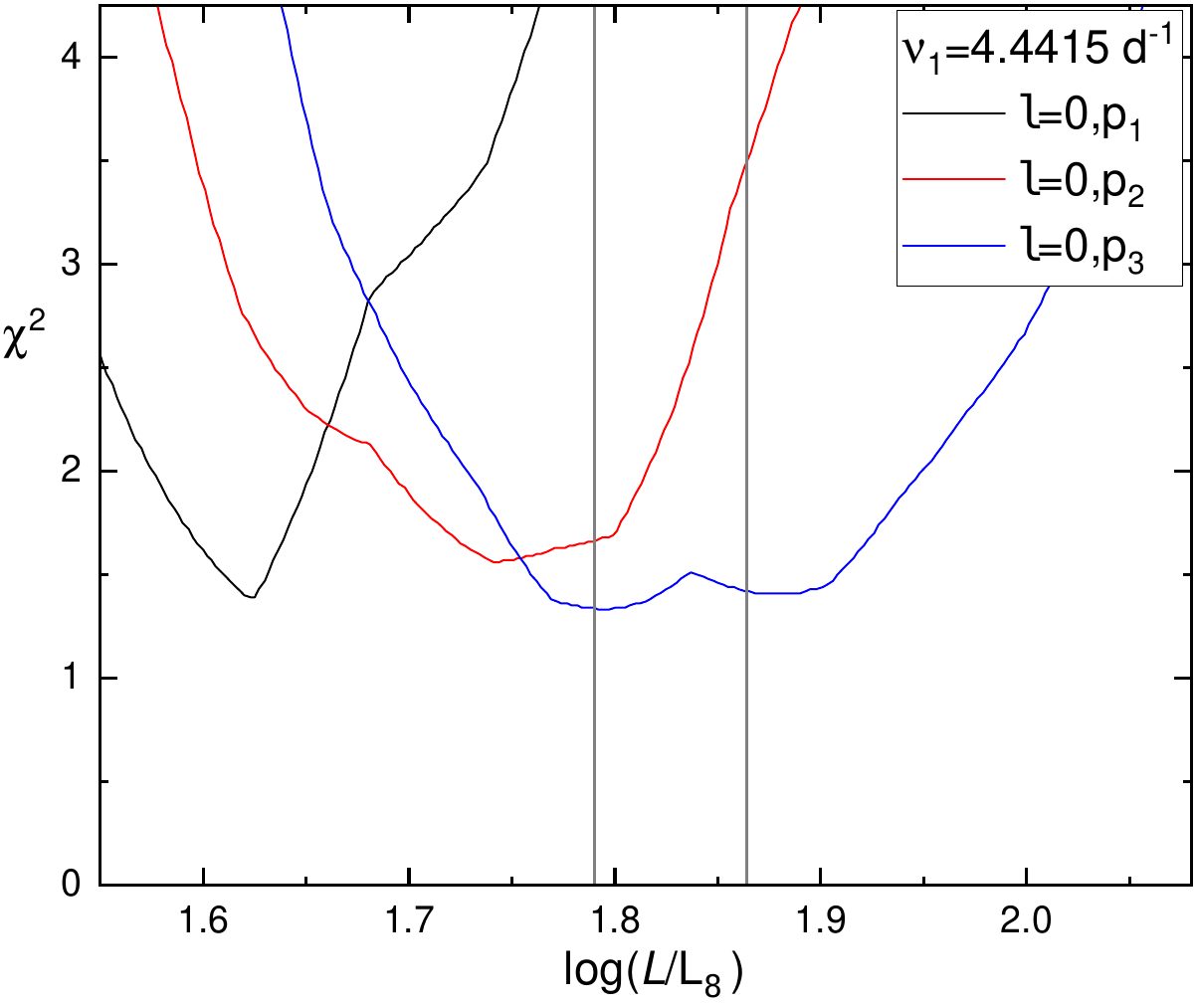}
	\caption{Left panel: the discriminant $\chi^2$ as a function of effective temperature for the first three radial modes.
		The parameters $(T_{\rm eff}, L)$ are taken from the lines of constant frequency 4.4415~d$^{-1}$. Model atmospheres 
		with [m/H] = -0.3 and $\xi_t=2$\,km\,s$^{-1}$ were adopted. Vertical lines indicate the range of $T_{\rm eff}$ derived from spectroscopy.
		Right panel: the same values of $\chi^2$ plotted as a function of the corresponding luminosities.}
	\label{fig:rylep_chi_lum}%
\end{figure*}

In the left panel of Fig.~\ref{fig:rylep_chi_lum}, we show $\chi^{2}$ as a function of the effective temperature for the first three radial modes. The parameters $(T_{\rm eff},\,L)$ are taken from the lines of constant frequency at $4.4415\,\mathrm{d}^{-1}$. The vertical lines indicate the observational range of $\log T_{\rm eff}$ for RY~Lep. All presented models fall within the observed effective-temperature range. The right panel of Fig.~\ref{fig:rylep_chi_lum} shows the same $\chi^{2}$ values but plotted as a function of the corresponding luminosities.The minima of $\chi^{2}$ for the first ($p_2$) and second radial overtones ($p_3$) lie within the observed luminosity range, whereas the minimum for the fundamental radial mode ($p_1$) falls outside the allowed range of luminosity. 
Thus, the dominant radial mode is either the first or the second overtone.

For the frequency $6.5991\,\mathrm{d}^{-1}$, the large uncertainties in the photometric amplitudes and phases prevent a reliable identification of the spherical degree $\ell$, and the remaining frequencies were not detected in the $uvby$ photometry.

\begin{deluxetable*}{lrr} 
	\centering
   	\tabletypesize{\normalsize}
	\tablewidth{\columnwidth}
	\tablecaption{The amplitudes and phases of the main frequency (4.4415\,d$^{-1}$) of RY~Lep derived from the $uvby$ photometry obtained between 1998 and 2002 at the Sierra Nevada Observatory. \label{tab:RY_Lep_uvby_Rodriguez}}
	\tablehead{
	\colhead{Passband} & \colhead{ Amplitude [mag]} & \colhead{ phase [rad] } }
	\startdata
	~~~~~~$u$ & 0.186(3)~~~~~  & 1.15(1)~~~\\ 
	~~~~~~$v$ & 0.245(3)~~~~~  & 1.06(1)~~~\\
	~~~~~~$b$ & 0.200(2)~~~~~  & 1.05(1)~~~\\
	~~~~~~$y$ & 0.160(2)~~~~~  & 1.03(1)~~~\\
	\hline
	\enddata
\end{deluxetable*}


\begin{figure*}
	\centering
	\includegraphics[width=0.9\textwidth]{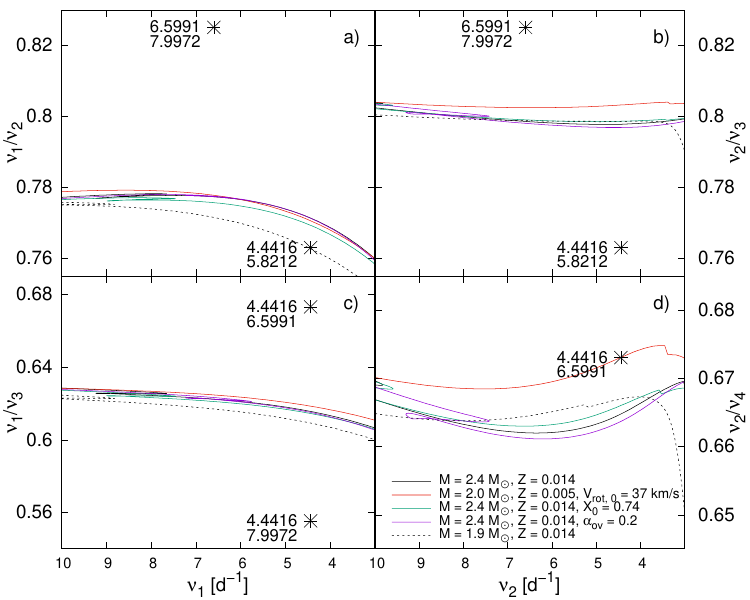}
	\caption{Petersen diagrams showing the frequency ratios for various pairs of radial modes as a function of the fundamental frequency (left panels) and the first-overtone frequency (right panels). The individual panels display the ratios of: (a) $p_1$ and $p_2$, (b) $p_2$ and $p_3$, (c) $p_1$ and $p_3$, and (d) $p_2$ and $p_4$. The effect of varying individual parameters is illustrated. The observed values for the selected frequency pairs are marked with asterisks.}%
	\label{fig:rylep_radial_pairs_tess}%
\end{figure*}

In the next step, we checked whether, in addition to the dominant frequency, another one could also correspond to a radial mode.
To this end, in Fig.~\ref{fig:rylep_radial_pairs_tess}, we plotted various types of Petersen diagrams. The four panels display the frequency ratios for different pairs of radial modes as a function of the fundamental frequency (left panels) or the first-overtone frequency (right panels). The observed values for various pairs of frequencies detected in the TESS light curves are also indicated. The theoretical computations illustrate the effects of mass, metallicity, initial hydrogen abundance, rotation, and the overshooting parameter.

The frequencies $4.4416\,\mathrm{d}^{-1}$ and $5.8212\,\mathrm{d}^{-1}$ yield a ratio typical of the fundamental and first-overtone radial modes, but the corresponding model parameters $(T_{\rm eff},\,L)$ are significantly lower than the observed values. For the pair $4.4416\,\mathrm{d}^{-1}$ and $6.5991\,\mathrm{d}^{-1}$, a possible identification is the $p_2$ and $p_4$ radial modes. The model has a metallicity of $Z=0.005$, and its $(T_{\rm eff},\,L)$ values lie within the observational error box. It is already in the HSB phase of evolution.

The remaining observed frequencies can only be associated with non-radial modes. Thus, the sole candidate for an additional radial mode is the frequency $6.5991\,\mathrm{d}^{-1}$.

\section{Bayesian seismic modeling}\label{sec:Seismic_modeling}

In our first approach to the seismic modeling of the primary pulsating component of the RY\,Lep system, 
we assume that its evolution can be approximated by that of a single star.

To compute stellar evolutionary models, we used the Warsaw–New Jersey code \citep[e.g.,][]{1999AcA....49..119P},
which assumes solid-body rotation and conservation of total angular momentum during evolution.
Convection in the stellar envelope is treated in the framework of standard mixing-length theory (MLT) and overshooting
from the convective core is implemented following the prescription of \citet{2008MNRAS.385.2061D}. 
This treatment accounts for both the overshooting distance $d_{\rm ov} = \alpha_{\rm ov}H_{\rm p}$,
where $H_{\rm p}$ is the pressure scale height and $\alpha_{\rm ov}$ is a free parameter, and for the hydrogen profile 
in the overshoot layer. We adopted OPAL opacity data \citep{1996ApJ...464..943I}, the AGSS09 chemical mixture and the OPAL2005 equation of state.

Stellar pulsations were computed using the linear nonadiabatic pulsation code of \citet{1977AcA....27...95D}. 
This code includes both first- and second-order rotational effects on the pulsation frequencies, treated within 
perturbation theory. It assumes that the convective flux remains unchanged during pulsation (the “frozen-in” approximation), 
which is justified when convection in the envelope is not highly efficient.

For RY~Lep, we achieved a unique mode identification only for the dominant frequency, which corresponds to a radial mode, 
most likely the first or second overtone, as shown in Sect.~\ref{sec:mode_identification}.
In our seismic analysis, we adopt two working hypotheses:\\
(1) the frequencies 4.4416\,d$^{-1}$ and 6.5991\,d$^{-1}$ correspond to the first and third overtone radial modes, respectively;\\
(2) the frequency 4.4416\,d$^{-1}$ is the first overtone radial mode, while 6.5991\,d$^{-1}$ is a dipole mode with an arbitrary azimuthal order $m$.\\
In addition to these two frequencies, we also fit the nonadiabatic parameter $f$ for the dominant mode. Its empirical values are derived from the photometric amplitudes and phases in the $uvby$ bands using the method of \citet{2003A&A...407..999D}, as described in Sect.~6.

To perform extensive seismic modeling of RY~Lep, we employed Bayesian inference based on Monte Carlo simulations.
Our goal was to reproduce the frequencies of two modes, the complex parameter $f$ of the dominant mode, and the stellar luminosity and effective temperature. The parameters to be determined were the mass $M$, metallicity $Z$, initial rotational velocity $V_{\rm rot,0}$, mixing-length parameter $\alpha_{\rm MLT}$, and overshooting parameter $\alpha_{\rm ov}$. Thus, we had six observables and five parameters to constrain.
For a convergent solution, the number of parameters to be determined cannot exceed the number of observables.
Median uncertainties were estimated from the 0.16 and 0.84 quantiles, corresponding to the standard deviations for a normal distribution.

The theoretical values of the nonadiabatic parameter $f$ were obtained from nonadiabatic pulsation computations.
In $\delta$~Scuti models, $f$ is highly sensitive to both the adopted opacity data and the efficiency of convection
in the outer layers \citep{2003A&A...407..999D,2023ApJ...942L..38D}. To derive the empirical values of $f$ from 
the $uvby$ amplitudes and phases, using the method of  \citet{2003A&A...407..999D}, we adopted 
the Vienna model atmospheres. In all simulations, we assumed a microturbulent velocity of $\xi = 4$\, km\,s$^{-1}$, as indicated by the analysis of the SALT spectrum, and an initial hydrogen abundance of $X_0 = 0.70$.

As noted earlier, we considered two possible identifications for the frequency 6.5991\,d$^{-1}$: either the radial third overtone or a dipole mode with three possible values of $m$. Consequently, we carried out four separate simulations. The median values of the parameter inferred from our seismic modeling are gathered in Table~\ref{tab:medians_all}. The first column gives the adopted $(\ell,m)$ identification for the two frequencies, followed by the corresponding radial orders $n$ in the second column. The third column gives the stellar mass and metallicity $Z$. The fourth column lists the initial rotation $V_{\rm rot, 0}$ and current rotation $V_{\rm rot}$. The fifth column contains the mixing-length parameter $\alpha_{\rm MLT}$ and the overshooting parameter $\alpha_{\rm ov}$. The subsequent column provides the model values of $(\log{T_{\rm eff}},\log(L/{\rm L_{\sun}})$ and the seventh column - the surface gravity $\log g$ and stellar radius $R$. The next column 
contains the age $t$ of seismic models and the rotation frequency $\nu_{\rm rot}$. The final two columns give the intrinsic mode amplitude $|\varepsilon|$ of the dominant frequency
and the instability parameter $\eta$ for both modes.

\begin{deluxetable*}{lllllllllllllll}
	\tabletypesize{\scriptsize}
\tablewidth{0pt}
			\tablecaption{Median values of the parameters of seismic models of RY~Lep in the HSB phase of evolution, obtained via Bayesian analysis based on Monte Carlo simulations. The models reproduce the frequency $\nu_1 = 4.4416$\,d$^{-1}$ as the first radial overtone and $\nu_2 = 6.5991$\,d$^{-1}$ either as the third radial overtone or as a dipole mode. A detailed description of all table columns is given in Sect.~7. 
 \label{tab:medians_all}}
				\tablehead{
                \colhead{modes} & \colhead{$n$ } & \colhead{ $M$ } & \colhead{ $V_{\rm rot, 0}$ } & \colhead{ $\alpha_{\rm MLT}$ } & \colhead{ $\log(T_{\rm eff}/{\rm K})$ } & \colhead{ $\log{g}$} & \colhead{ $t$  } & \colhead{ ~~$|\varepsilon|$ } & \colhead{ ~~$\eta$}
                \\[-9pt]
                \colhead{($\ell_1,m_1$)} & \colhead{ } & \colhead{ [M$_{\sun}$] } & \colhead{  [km\,s$^{-1}$] } & \colhead{ } & \colhead{ } & \colhead{ [cgs] } & \colhead{[Myr]} & \colhead{ } & \colhead{}
                \\[-5pt]
                \colhead{($\ell_2,m_2$)} & \colhead{ } & \colhead{ $Z$ } & \colhead{  $V_{\rm rot}$ } & \colhead{ $\alpha_{\rm ov}$ } & \colhead{  $\log(L/{\rm L_{\sun}})$ } & \colhead{ $R$ } & \colhead{ $\nu_{\rm rot}$} & \colhead{ } & \colhead{}
                \\[-9pt]
                \colhead{ } & \colhead{ } & \colhead{ } & \colhead{ [km\,s$^{-1}$] } & \colhead{  } & \colhead{  } & \colhead{ [R$_{\sun}$] } & \colhead{ [d$^{-1}$] } & \colhead{ } & \colhead{}
                }
                \startdata
        (0,0)& $p_2$ & $2.006^{+0.049}_{-0.058}$ & $82.1^{+13.9}_{-35.1}$ & $0.26^{+0.29}_{-0.18}$ & $3.8598^{+0.0070}_{-0.0080}$ & $3.264^{+0.006}_{-0.003}$  & $785^{+86}_{-132}$ & $0.0101^{+0.0009}_{-0.0008}$ &  $0.028^{+0.035}_{-0.040}$ \\
		(0,0)& $p_4$ & $0.0054^{+0.0014}_{-0.0013}$ &    $59.7^{+13.2}_{-25.2}$ &    $0.38^{+0.09}_{-0.14}$ &    $1.852^{+0.030}_{-0.034}$ &    $5.372^{+0.049}_{-0.038}$ & $0.217^{+0.051}_{-0.101}$  & &  $0.121^{+0.013}_{-0.017}$  \\
		\hline
        \hline
		\vspace{0.1cm}
		(0,0)& $p_2$   &  $2.131^{+0.017}_{-0.097}$ &    $14.9^{+2.4}_{-2.5}$ &    $0.28^{+0.30}_{-0.19}$ &    $3.8536^{+0.0078}_{-0.0090}$ &    $3.285^{+0.002}_{-0.008}$ &    $743^{+61}_{-65}$ & $0.0087^{+0.0006}_{-0.0003}$ &  $0.042^{+0.032}_{-0.041}$ \\
		(1,-1)  &$g_{30}$ - $g_{37}$ &    $0.0085^{+0.0020}_{-0.0024}$ &    $10.8^{+1.8}_{-1.8}$ &    $0.32^{+0.09}_{-0.04}$ &    $1.849^{+0.028}_{-0.044}$ &    $5.503^{+0.023}_{-0.075}$ & $0.039^{+0.006}_{-0.006}$ & &   $0.125^{+0.020}_{-0.015}$  \\ 
		\hline
		\vspace{0.1cm}
		(0,0)& $p_2$  &  $2.190^{+0.093}_{-0.130}$ &    $20.6^{+16.4}_{-14.5}$ &    $0.27^{+0.32}_{-0.18}$ &    $3.8548^{+0.0080}_{-0.0089}$ &    $3.288^{+0.007}_{-0.011}$ &    $630^{+137}_{-123}$ &  $0.0087^{+0.0009}_{-0.0003}$ & $0.033^{+0.032}_{-0.036}$\\
		(1,0) &    $g_{33}$ - $g_{51}$ &    $0.0080^{+0.0017}_{-0.0019}$ &    $14.1^{+13.1}_{-10.1}$ &    $0.22^{+0.13}_{-0.14}$ &    $1.860^{+0.033}_{-0.040}$ &    $5.560^{+0.065}_{-0.130}$ & $0.051^{+0.048}_{-0.037}$ & &   $0.124^{+0.015}_{-0.013}$ \\ 
		\hline
		\vspace{0.1cm}
		(0,0)& $p_2$& $2.096^{+0.069}_{-0.018}$ &    $21.7^{+16.3}_{-8.5}$ &    $0.27^{+0.29}_{-0.19}$ &    $3.8531^{+0.0100}_{-0.0041}$ &    $3.282^{+0.004}_{-0.002}$  &    $813^{+78}_{-251}$ & $0.0087^{+0.0010}_{-0.0002}$ & $0.040^{+0.024}_{-0.040}$ \\
		(1,+1)  &    $g_{25}$-$g_{52}$ &    $0.0084^{+0.0013}_{-0.0028}$ &    $15.9^{+10.2}_{-5.9}$ &    $0.41^{+0.08}_{-0.28}$ &    $1.840^{+0.041}_{-0.017}$ &    $5.468^{+0.062}_{-0.009}$ & $0.057^{+0.036}_{-0.021}$ & &  $0.124^{+0.018}_{-0.013}$  \\
	\hline
	\hline
	\multicolumn{7}{l}{Two radial mode hypothesis after selection for $\nu_{\rm rot} \in (0.157, 0.160)$:}\\
	\hline
		\vspace{0.1cm}
		(0,0)& $p_2$   &	 $1.998^{+0.041}_{-0.048}$ &    $61.8^{+2.8}_{-3.6}$ &    $0.23^{+0.25}_{-0.15}$ &    $3.8600^{+0.0058}_{-0.0063}$ &    $3.268^{+0.003}_{-0.004}$  &    $703^{+88}_{-68}$& $0.0109^{+0.0003}_{-0.0004}$ &   $0.025^{+0.028}_{-0.030}$ \\
		(0,0)& $p_4$ &   $0.0044^{+0.0002}_{-0.0002}$ &    $43.2^{+0.4}_{-0.4}$ &    $0.29^{+0.11}_{-0.08}$ &    $1.856^{+0.020}_{-0.026}$ &    $5.387^{+0.035}_{-0.039}$ &    $0.158^{+0.001}_{-0.001}$ & &    $0.121^{+0.009}_{-0.011}$ \\
\enddata
\end{deluxetable*}

Histograms of $M$, $Z$, age, the current rotational velocity $V_{\rm rot}$, $\alpha_{\rm MLT}$, and $\alpha_{\rm ov}$
from all four simulations are presented in Appendix~\ref{sec:appendix_histograms}.

\subsection{Seismic models from two-radial mode hypothesis}
In the case of the two-radial-mode hypothesis, the median seismic mass of RY~Lep is $M \approx 2.006\,{\rm M}_{\sun}$.
The metallicity is $Z \approx 0.005$, corresponding to [m/H] $\approx -0.5$, in agreement with spectroscopic determinations. 
Notably, this metallicity is significantly lower than the typical value observed for Galactic disk stars of comparable age \citep{1998MNRAS.296.1045C}. The seismic age of RY~Lep lies in the range 653–871 Myr.

The median initial rotational velocity is $\sim 80$\,km\,s$^{-1}$, with a current value $\sim 60$\,km\,s$^{-1}$.  Given the observed value $V_{\rm rot}\sin{i} = 21.9$\,km\,s$^{-1}$ and the rotation range from our simulations (34.5–72.9)\,km\,s$^{-1}$, the inclination angle of the pulsating component of the RY~Lep system is constrained to $i \in (17^\circ, 40^\circ)$. The range of rotation frequency from our simulations is $\nu_{\rm rot} \in (0.116, 0.268)$\,d$^{-1}$.  The low frequency signal, $\nu_{\rm low}\approx 0.16$\,d$^{-1}$, detected in the TESS data falls within this range.

The median value of the mixing-length parameter inferred from our simulations is $\alpha_{\rm MLT} \in (0.08,~0.55)$,
indicating rather inefficient convection in the outer stellar layers. In contrast, the overshooting parameter is relatively large,
$\alpha_{\rm ov} \in (0.24, 0.47)$, which may reflect compensation for various mixing processes not included in the Warsaw–New Jersey code. Both frequencies, 4.4416\,d$^{-1}$ and 6.5991\,d$^{-1}$, have positive values of the instability parameter $\eta$, indicating that these modes are excited in our seismic models.

For the dominant frequency, it was also possible to determine the empirical value of the intrinsic mode amplitude $\varepsilon$,
because the photometric amplitudes of radial modes are independent of the inclination angle $i$. The absolute value $|\varepsilon|$
represents the root-mean-square (RMS) of $\delta r / R$ over the stellar surface. Although $\varepsilon$ has no direct counterpart in linear theory, it provides an estimate of the stellar radius variations. For the dominant mode of RY~Lep, we obtained $|\varepsilon| \approx 1\%$.

We further selected a set of seismic models that reproduce the observed low-frequency signal, $\nu_{\rm low} \in (0.157, 0.160),\mathrm{d}^{-1}$, 
interpreting it as the stellar rotation frequency. These models were constrained to have a metallicity of $Z \approx 0.004$, 
a radius of $R \approx 5.4\,R_{\odot}$, and a surface rotational velocity of $V_{\rm rot} \approx 43\,\mathrm{km,s}^{-1}$.
Their effective temperature, $\log(T_{\rm eff}/\mathrm{K}) \approx 3.86$, is slightly higher than the range inferred from spectroscopy.

\subsection{Seismic models from radial and dipole mode hypothesis}
This time, we assume that the frequency 4.4416\,d$^{-1}$ corresponds to the first radial overtone, while 6.5991\,d$^{-1}$ is a dipole mode ($\ell = 1$) with an unknown azimuthal order, $m = -1, 0,$ or $+1$.
In all cases, the 6.5991\,d$^{-1}$ frequency is reproduced by high-order gravity ($g$) modes with radial orders $n \ge 25$.

\begin{figure*}
	\centering
	\includegraphics[width=0.98\textwidth, height=9cm]{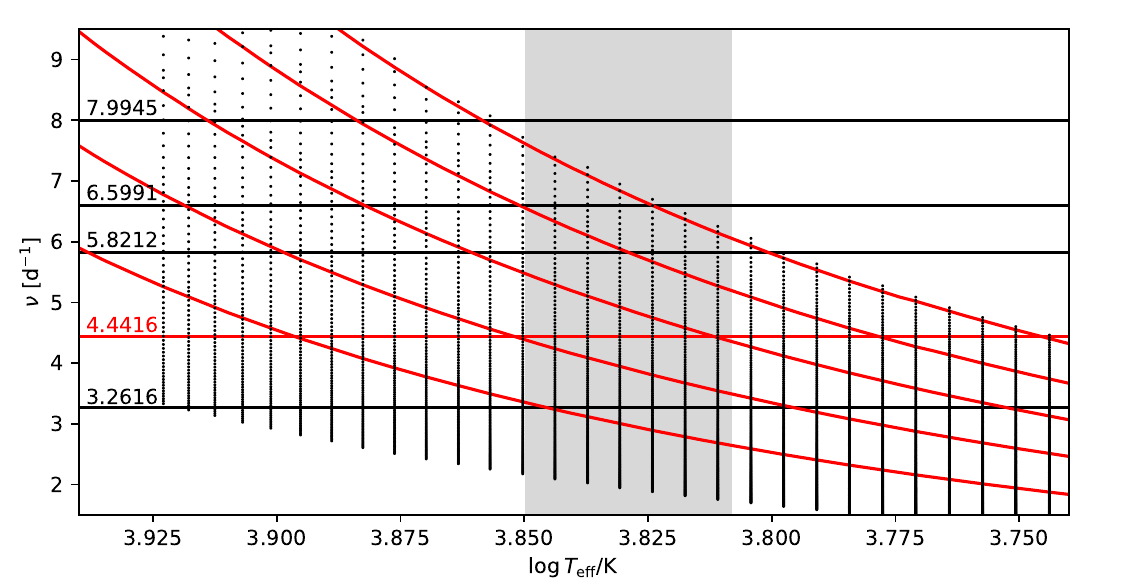}
	\caption{Evolution of the frequencies of radial and dipole modes during the HSB phase, shown as a function of effective temperature, for a model with $M = 2.4\,{\rm M}_{\sun}$, $X_0 = 0.7$, $Z = 0.008$, $V{\rm rot,0} = 45$\,km\,s$^{-1}$, $\alpha_{\rm MLT} = 0.5$, and $\alpha_{\rm ov} = 0$. Red lines denote radial modes (the lowest line corresponds to $p_1$), and the small black dots show dipole axisymmetric modes ($\ell = 1$, $m = 0$). Horizontal lines mark the frequencies observed in the TESS light curve; the dominant frequency at 4.4416\,d$^{-1}$ is shown in red. The shaded region marks the range of effective temperatures we derived from spectroscopy.}%
	\label{fig:rylep_dipole_modes}%
\end{figure*}

\begin{figure}
	\centering
	\includegraphics[width=0.45\textwidth]{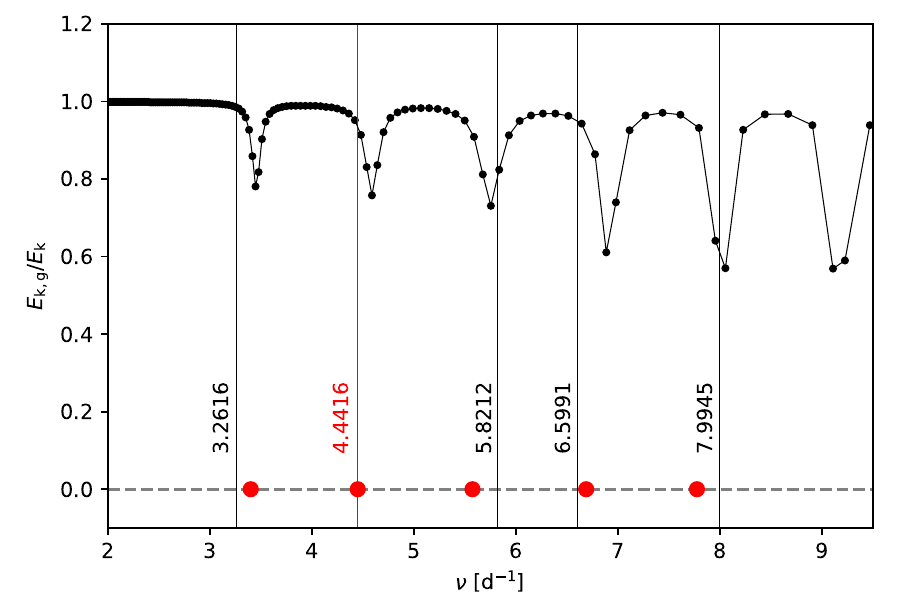}
	\caption{Ratio of the kinetic energy in the gravity-wave propagation zone to the total kinetic energy, $E_{\rm k,g}/E_{\rm k}$, for dipole axisymmetric modes as a function of frequency. The model shown reproduces the frequency 4.4416\,d$^{-1}$ as the first radial overtone. Red dots mark the frequencies of radial modes and vertical lines indicate the observed frequencies derived from the TESS data.}%
	\label{fig:rylep_ekg}%
\end{figure}

In Fig.~\ref{fig:rylep_dipole_modes}, we present the evolution of the frequencies of radial and dipole modes during the HSB phase for a model with $M=2.4\,{\rm M}_{\sun}$, whose evolutionary track is shown in Fig.~\ref{fig:rylep_hr_diagram}. 
The spectrum of nonradial modes becomes very dense in the HSB phase. In these models, the Brunt–Väisälä frequency is very large in the deep interior, with its maximum located at the boundary of the small helium core. As a result, the kinetic energy density of nonradial modes is large and strongly concentrated within the helium core \citep{2023MNRAS.526.1951D}. Consequently, all nonradial modes acquire a dual character with a strong gravity-mode component.
This is illustrated in Fig.~\ref{fig:rylep_ekg}, where we show the ratio of kinetic energy in the gravity-wave propagation zone to the total kinetic energy, $E_{\rm k,g}/E_{\rm k}$, for the dipole modes in the model that reproduces the frequency 4.4416\,d$^{-1}$ as the first radial overtone (cf. Fig.~\ref{fig:rylep_dipole_modes}). The dipole modes near the frequencies 3.2616, 5.8212, and 6.5991\,d$^{-1}$ have kinetic 
energies dominated by their gravity-mode components. Modes near the highest frequency 7.9945\,d$^{-1}$ have the ratio $E_{\rm k,g}/E_{\rm k}$ 
of about 60\%. Minima in $E_{\rm k,g}/E_{\rm k}$ for the dipole modes occur near the frequencies of radial modes, which are indicated 
by the red dots.

The seismic models yield a stellar mass of $\sim 2.1$\,M$_{\sun}$ and a metallicity of $Z \approx 0.008$ ([m/H]$\approx -0.3$), 
with ages in the range 500–900\,Myr. 
As in the two-radial-mode scenario, convection in the outer layers is inefficient, with a median mixing-length parameter of $\alpha_{\rm MLT} \approx 0.3$. The inferred extent of convective-core overshooting depends on the azimuthal order: for $m=-1$, $\alpha_{\rm ov} = 0.32^{+0.04}_{-0.09}$; for $m=0$, $\alpha_{\rm ov}$ is poorly constrained; and for $m=+1$, most models reach the simulation upper limit, $\alpha_{\rm ov} \approx 0.5$ (cf. Fig.~\ref{fig:rylep_l0_l1_mp1_histogram}), indicating similarly weak constraints.

The current rotational velocity is in the range $V{\rm rot} \in (4,~27)$\,km\,s$^{-1}$ and most seismic models
have rotation below the observed $V_{\rm rot}\sin{i} = 21.9$\,km\,s$^{-1}$, although pulsations may also contribute to the broadening of spectral lines. 
However, in all three cases ($m = -1, 0, +1$), the rotation frequencies is significantly smaller than the low-frequency peak $\nu_{\rm low}\approx 0.16$\,d$^{-1}$.  Thus, if $\nu_{\rm low}$ corresponds to the rotation frequency then the radial-dipole-mode hypothesis must be rejected.

We also considered the alternative in which the dominant frequency corresponds the second radial overtone, with 6.5991\,d$^{-1}$ still representing a dipole mode. This scenario is supported by the constant-frequency line within the error box in the HR diagram (Fig.~\ref{fig:rylep_hr_diagram}) and by the behaviour of the $\chi^2$ discriminant (Fig.~\ref{fig:rylep_chi_lum}).
However, these models consistently predict a surface gravity of $\log{g} \approx 3.15$, which is significantly lower
than photometric and spectroscopic estimates.

\section{Discussion}\label{sec:discussion}

As noted in Sect. 2, RY~Lep was also classified as an RR Lyrae star based on its pulsation periods and amplitudes, 
which fall within the range defined by \citet{2023A&A...674A..18C} from Gaia data. However, its surface gravity,
derived from photometric and spectroscopic analyses, exceeds the values for metal-rich RR Lyrae stars, 
for which $\log{g} < 2.6$ \citep{2017ApJ...835..187C}. Additionally, RY~Lep’s iron abundance, [Fe/H] = –0.44(1), 
is significantly higher than the typical values for RRc-type variables, which are usually $< -1.0$ \citep{2025MNRAS.536.2749M}. Furthermore, RR Lyrae stars are typically low-mass objects ($M < 0.8$\,M$_{\sun}$; \citealt{1993AJ....106..703S}),
not consistent with RY~Lep's mass. Given these discrepancies, the most reliable classification for RY~Lep
is as a High Amplitude $\delta$ Scuti star.

Our spectral analysis reveals a relatively low calcium abundance ([Ca/H] = –0.39(5)) accompanied by overabundances 
of yttrium and barium, which could suggest an Fm-type classification, potentially resulting from atomic diffusion
driven by radiative acceleration \citep{2000ApJ...529..338R}. However, the scandium and iron-group element abundances
differ from typical AmFm stars, and the calcium-to-scandium ratio ([Ca/Sc] = –0.28) deviates from the usual AmFm pattern,
which typically exceeds +0.6 \citep{2024MNRAS.527.8234M}.
A similar heavy-element pattern is observed in $\rho$~Puppis stars, where barium and europium are overabundant \citep{2020AJ....160...52M}, potentially indicating mass transfer from an AGB companion via Roche-lobe overflow (Case~C or D; \citealt{2007ASPC..372..397M}). The elevated abundance of $r$-process elements ($Z > 60$) further suggests a contribution from $r$-process nucleosynthesis.

The simultaneous enhancement of both $s$- and $r$-process elements can be explained by the intermediate neutron-capture ($i$-) process. While the $i$-process is typically associated with low metallicities ([Fe/H] $< -2.0$; \citealt{2022A&A...667A.155C}),
in 2\,M$_{\odot}$ stars with [Fe/H] $\approx -0.5$, convective overshooting may trigger proton-ingestion events (PIEs),
producing an $i+s$ signature at the AGB surface \citep{2024A&A...684A.206C}.

As shown in Fig.~\ref{fig:rylep_sed}, RY~Lep exhibits excess far-UV emission, which could arise from a white dwarf companion. 
An alternative explanation might involve chromospheric activity, as suggested for some $\delta$~Scuti stars, for example $\beta$~Cas \citep{1989ApJ...343..916T,1991ApJ...375..704A,2003AdSpR..31..381D}. However, there is no direct spectroscopic evidence for chromospheric emission in RY~Lep.
Our asteroseismic Monte Carlo simulations yield a median mass of $\sim 2$\,M$_{\sun}$ for RY~Lep. Assuming a companion mass
of $0.51^{+0.14}_{-0.06}$\,M$_{\sun}$ \citep{2019A&A...623A..72K}, the minimum initial system mass would be $\sim 2.5$\,M$_{\sun}$. If both stars initially
had masses of $\sim 1.25$\,M$_{\sun}$, mass transfer must have occurred roughly 4\,Myr after formation, with substantial mass
loss likely during AGB thermal pulses.

The effects of binary evolution, including mass transfer under Case~C and D scenarios and their impact on pulsation properties, will be examined in a separate study.

\section{Summary}\label{sec:summary}

We performed a comprehensive analysis of RY~Lep, a relatively bright HADS star.
Radial velocity measurements, phase shifts in the light curves, and perturbations in proper motion indicate that RY~Lep
is part of a binary system, likely hosting a white dwarf companion.

Spectroscopic analysis revealed a significantly lower metallicity than previously reported, along with an overabundance
of heavy elements produced via neutron-capture processes. These chemical signatures may indicate past mass transfer from
a former asymptotic giant branch (AGB) companion, likely through Case~C or D mass transfer. Based on its position
in the HR diagram, the primary component of RY~Lep is classified as a post-main-sequence star.
Taken together with the evidence for binarity, RY~Lep could be identified as a yellow straggler.

Analysis of TESS photometric data from sectors S06, S32, and S33 revealed several previously unreported secondary frequencies, all exhibiting strongly modulated amplitudes. The previously reported frequency at 6.6\,d$^{-1}$ was detected only in sector S06 as well as 
in the ASAS and SuperWASP light curves. 
Multiband $uvby$ photometry allowed us to identify the dominant frequency of 4.4415\,d$^{-1}$ as a radial mode, most likely the first overtone. The secondary frequency at 6.6\,d$^{-1}$ may correspond either to a radial or a nonradial mode, while all other secondary frequencies can only be nonradial modes.

Seismic modeling based on MC simulations confirmed that RY~Lep is a post-main sequence star, with all models in the HSB phase of evolution. Two hypotheses were considered for seismic modeling: (1) both 4.44\,d$^{-1}$ and 6.6\,d$^{-1}$ correspond to radial modes (first and third overtones, respectively), and (2) 4.44\,d$^{-1}$ is the first overtone radial mode while 6.6\,d$^{-1}$ is a dipole mode ($\ell = 1$) with azimuthal orders $m = -1, 0, +1$, chosen as the mode with the best visibility after the radial mode.
In both scenarios, the models yield small values for the mixing length parameter, $\alpha_{\rm MLT} \in (0.1, 0.5)$, indicating that convective energy transport in the outer layers is relatively inefficient. 
The low-frequency peak, $\nu_{\rm low} \approx 0.16$\,d$^{-1}$, detected in the TESS data, may correspond to the rotation frequency, which imposes additional strong constraints on the seismic models. If this is the case, then only the two-radial-mode hypothesis
for the RY~Lep pulsator is viable. 

Many models also show a substantial extent of overshooting from the convective core, $\alpha_{\rm ov} \in (0.1, 0.5)$. This may reflect limitations of the Warsaw-New Jersey code, which does not account for additional mixing processes such as those induced by rotational instabilities, or it may result from the simplifications of using a single evolutionary code for seismic modeling.
Further modelling using more advanced codes that incorporate binary evolution and additional mixing processes is required to fully characterize RY~Lep. In particular, this system offers an excellent opportunity to test theories of mass transfer from AGB companions, a topic that will be addressed in our future study.

\begin{acknowledgments}
The work was financially supported by the Polish National Science Centre grant MAESTRO 2023/50/A/ST9/00144.
ER acknowledges financial support from the Spanish Agencia Estatal de Investigaci\'on (AEI/10.13039/501100011033) of the Ministerio de Ciencia e Innovaci\'on through projects PID2022-137241NB-C43 and PID2023-149439NB-C42. PKS was supported by the University of Liège under the Special Funds for Research, IPD-STEMA Programme, and by the National Science Centre, Poland, grant no. 2022/45/B/ST9/03862.

This work has made use of data from the European Space Agency
(ESA) mission Gaia (https://www.cosmos.esa.int/gaia), processed
by the Gaia Data Processing and Analysis Consortium (DPAC; https:
//www.cosmos.esa.int/web/gaia/dpac/consortium). Funding for the
DPAC has been provided by national institutions, in particular the
institutions participating in the Gaia Multilateral Agreement.

This paper includes data collected by the TESS mission. Funding
for TESS is provided by the NASA Explorer Program.
\end{acknowledgments}

\section*{Data Availability}
The TESS light curves were downloaded from the TASOC website https://tasoc.dk.
The other photometric and spectroscopic data are available in the literature or databases.
Theoretical computations of stellar evolution and pulsations will be shared 
on reasonable request to the corresponding author.

\facilities{Gaia, TESS, SuperWASP, ASAS, OSN:0.9m, ATT, SALT}
\software{
iSpec \citep{2014A&A...569A.111B,2019MNRAS.486.2075B},
Spectrum \citep{1994AJ....107..742G},
astropy \citep{2013A&A...558A..33A,2018AJ....156..123A},
numpy \citep{2020Natur.585..357H},
matplotlib \citep{2007CSE.....9...90H}
}

\newpage

\appendix
\restartappendixnumbering

\section{The SALT spectrum of RY Lep}\label{sec:appendix_spectrum}

\begin{figure*}
    \centering
    \includegraphics[width=0.95\textwidth]{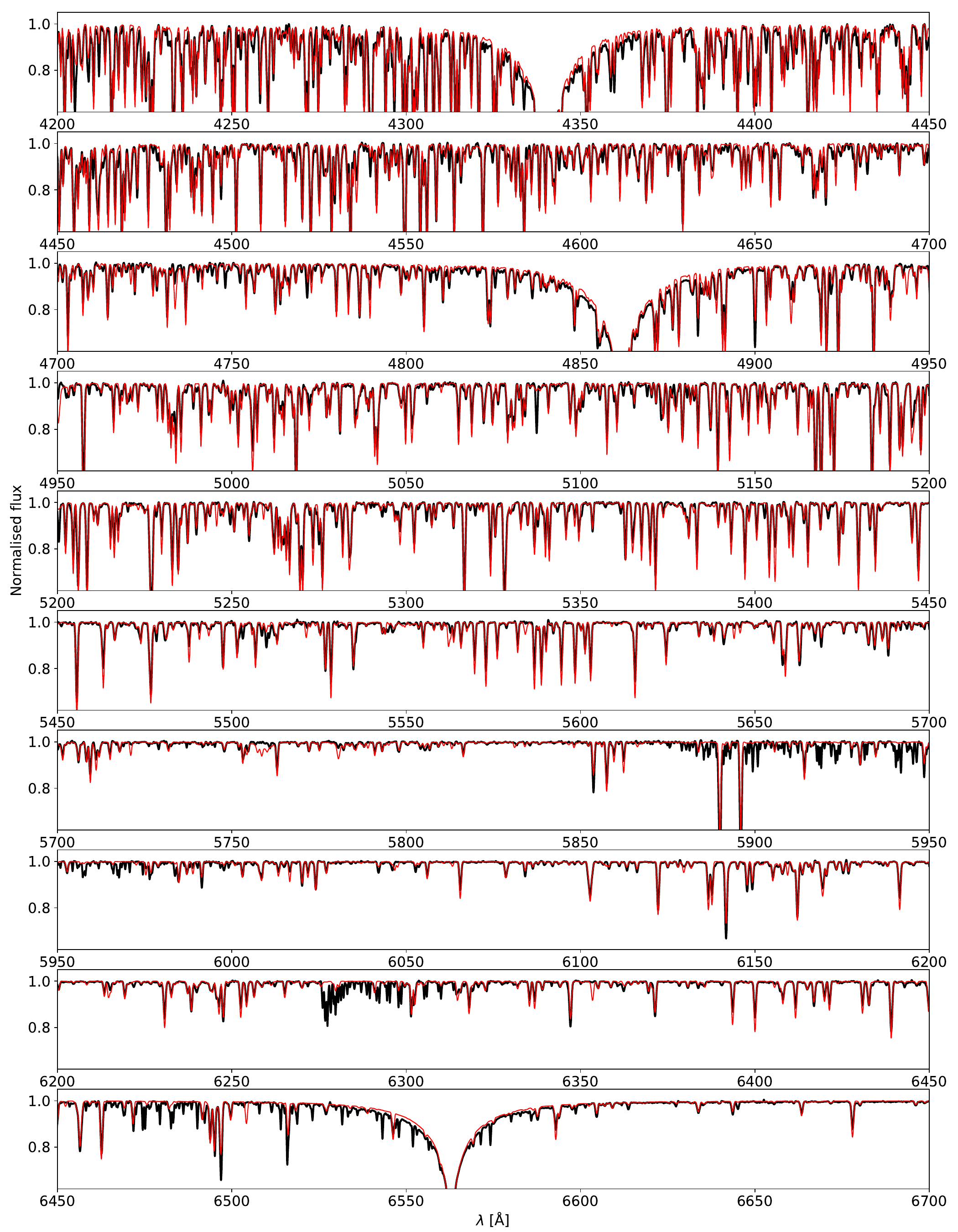}
    \caption{The black line shows the SALT spectrum of RY~Lep over the entire observed wavelength range. The red line indicates the synthetic
     spectrum computed for $T_{\rm eff}=6750$\,K, $\log{g}=3.5$, [m/H] = -0.4, $\xi=4$\,km\,s$^{-1}$ and $V_{\rm rot}\sin{i} = 20$\,km\,s$^{-1}$.}%
    \label{fig:rylep_spectrum_SALT}%
\end{figure*}

\begin{figure*}
	\centering
	\includegraphics[width=\textwidth]{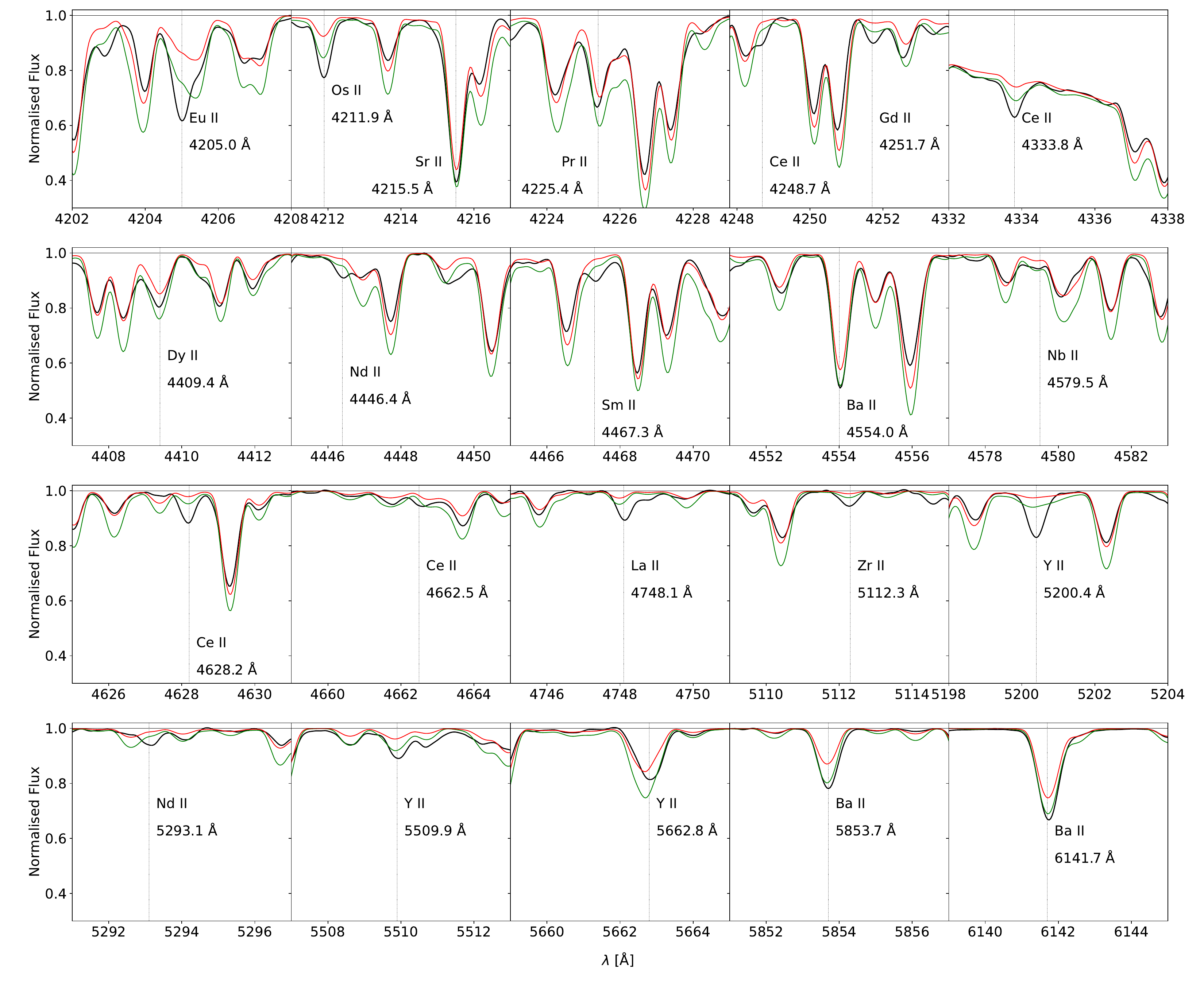}
	\caption{Spectral lines of chemical elements with $Z>30$ showing overabundance, as observed in the SALT spectrum of RY~Lep (black lines). The synthetic spectra were computed for	$T_{\rm eff}=6878$\,K, $\log{g}=3.51$, $V_{\rm rot}\sin{i} = 21.9$\,km\,s$^{-1}$, $\xi=4$\,km\,s$^{-1}$ and two pairs of the metallicity [m/H] and abundance of $\alpha-$elements $[\alpha/\rm Fe]$. The red line corresponds to [m/H]$ = -0.4$, [$\alpha/\rm Fe] = 0.1$, and the green line to [m/H] = 0.0, [$\alpha/\rm Fe] = 0.0$.}%
	\label{fig:rylep_spectrum_SALT_lines}%
\end{figure*}

In Fig.~\ref{fig:rylep_spectrum_SALT}, we present the SALT spectrum of RY~Lep over the wavelength range used in our analysis. The lines of significant overabundant elements with $Z>30$ are depicted in Fig.~\ref{fig:rylep_spectrum_SALT_lines}.

\restartappendixnumbering

\section{Pulsation frequencies of RY Lep from various observations}\label{sec:appendix_frequencies}

\begin{figure}
	\centering
	\includegraphics[width=0.45\textwidth]{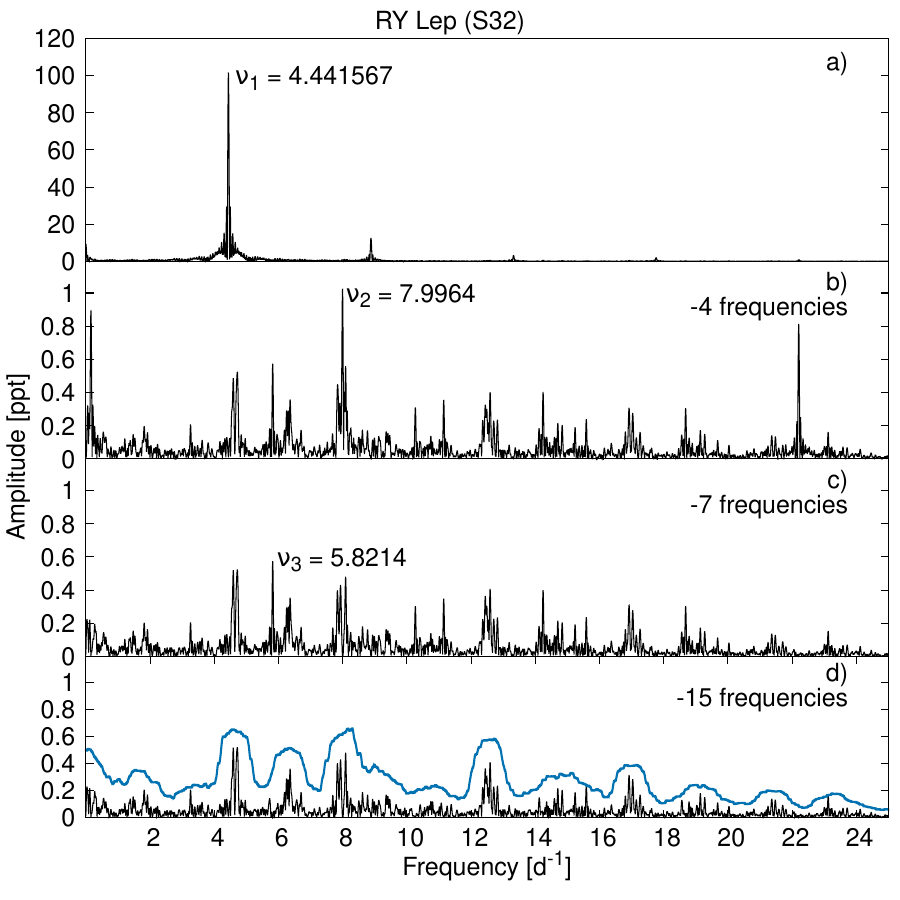}
	\caption{Fourier amplitude periodograms of RY~Lep derived from the TESS light curve in Sector S32. From top to bottom, the panels show: the periodogram for the original data, after subtracting four terms, after subtracting 7 terms and after subtracting 15 terms. The blue horizontal line marks the signal-to-noise (S/N) threshold of 5.}%
	\label{fig:periodogram_TESS_S32}%
\end{figure}

\begin{figure}
	\centering
	\includegraphics[width=0.45\textwidth]{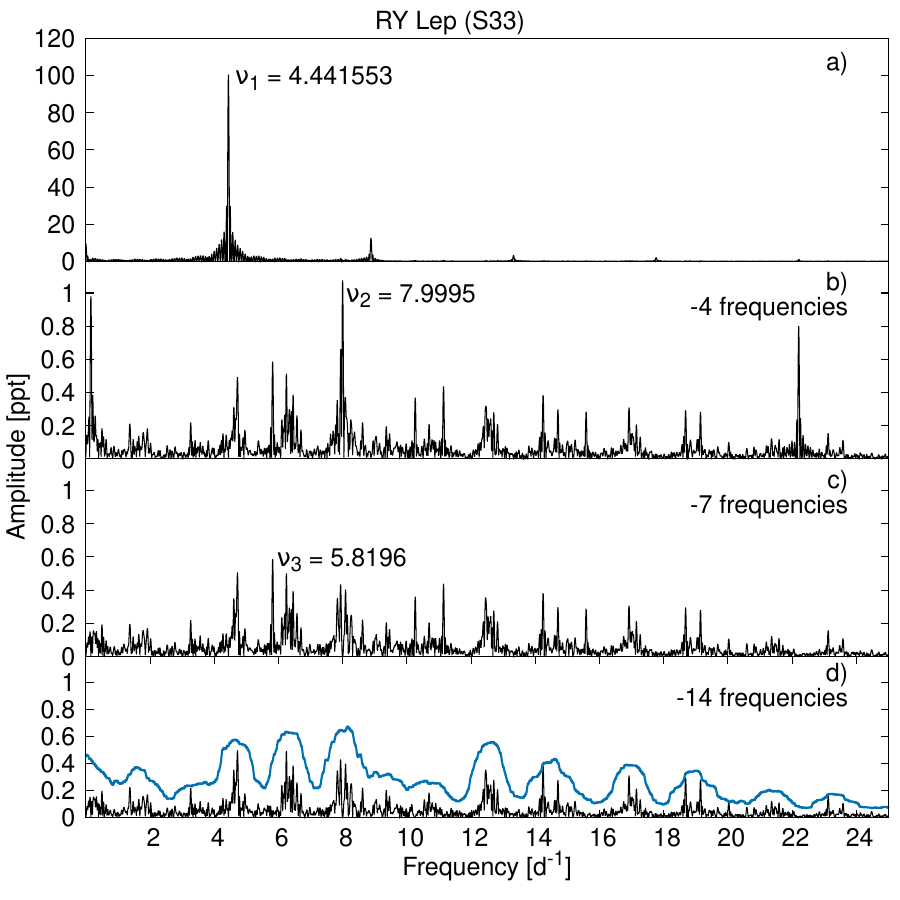}
	\caption{Fourier amplitude periodograms of RY~Lep derived from the TESS light curve in Sector S33. From top to bottom, the panels show: the periodogram for the original data, after subtracting four terms, after subtracting 7 terms and after subtracting 14 terms. The blue horizontal line marks the signal-to-noise (S/N) threshold of 5.}%
	\label{fig:periodogram_TESS_S33}%
\end{figure}

In this section, we list the frequencies of RY~Lep extracted from three photometric datasets. 
Frequencies found in TESS Sector S06 and the combined data of Sectors S32 and S33 
are listed in Tables~\ref{freq_RY_Lep_S06} and~\ref{freq_RY_Lep_S32_S33}, respectively.
Tables~\ref{freq_RY_Lep_S32} and~\ref{freq_RY_Lep_S33} contain frequencies obtained separately from S32 and S33, respectively. 
Figures~\ref{fig:periodogram_TESS_S32} and~\ref{fig:periodogram_TESS_S33} show the Fourier amplitude periodograms for the original data and
after three pre-whitening steps for Sectors S32 and S33, respectively.

In Table~\ref{freq_RY_Lep_ASAS}, we show the frequencies extracted from ASAS data, both for the full 9-year dataset and for shorter subdivided time segments. Table~\ref{freq_RY_Lep_superwasp}  lists the frequencies identified in the SuperWASP data, divided into three time segments.
\begin{deluxetable}{|l|r|r|r|l|}[t]
\tabletypesize{\scriptsize}
			\tablecaption{The complete set of frequencies found in the TESS observations of RY~Lep from Sector S06. \label{freq_RY_Lep_S06}}
	\tablehead{
                \colhead{ } & \colhead{  Frequency } & \colhead{  A } & \colhead{  S/N } & \colhead{  ID} \\
		  \colhead{ } & \colhead{  $[\rm d^{-1}]$ } & \colhead{  [ppt] } & \colhead{  } & \colhead{}
        }
        \startdata 
        \hline
		  1  & 4.441626(4)  &  104.21(2)   &  8.46  &  $\nu_1$   \\ 
		  2  &  8.88323(3)  &   15.97(2)   &  8.46  &  $2\nu_1$   \\ 
		  3  &  13.3250(1)  &    3.93(2)   &  8.50  &  $3\nu_1$   \\ 
		  4  &  17.7667(2)  &    2.25(2)   &  8.46  &  $4\nu_1$   \\ 
		  5  &   6.5991(3)  &    1.74(2)   &  6.41  &  $\nu_2$   \\ 
		  6  &  22.2083(5)  &    0.95(2)   &  8.37  &  $5\nu_1$   \\ 
		  7  &  11.0399(5)  &    0.83(2)   &  6.45  &  $\nu_1+\nu_2$   \\ 
		  8  &   7.9945(6)  &    0.81(2)   &  5.72  &  $\nu_3$   \\ 
		  9  &  26.6498(9)  &    0.51(2)   &  8.08  &  $6\nu_1$   \\ 
		 10  &   31.092(2)  &    0.27(2)   &  7.67  &  $7\nu_1$   \\ 
		 11  &   35.535(3)  &    0.13(2)   &  7.08  &  $8\nu_1$   \\ 
		 12  &   39.976(7)  &    0.07(2)   &  5.38  &  $9\nu_1$   \\ 
		\hline
        \enddata
\end{deluxetable}
\begin{deluxetable}{|l|r|r|r|l|}[t]
\tabletypesize{\scriptsize}
\tablewidth{0pt}
		\tablecaption{The frequencies found in the TESS light curves of RY~Lep from the combined  S32+S33 sectors. Bold comments indicate frequencies for which the difference between observed frequency and possible combination is between $1/T$ and $2.5/T$. \label{freq_RY_Lep_S32_S33}}
	\tablehead{
        \colhead{ } & \colhead{  Frequency } &    \colhead{  A } & \colhead{  S/N } & \colhead{  ID} \\
		  \colhead{ } & \colhead{  $[\rm d^{-1}]$ } & \colhead{  [ppt] } & \colhead{  } & \colhead{}
        }
        \startdata
        \hline
		  1  & 4.441552(1)  &  100.71(1)   & 15.51  &  $\nu_1$   \\ 
		  2  & 8.883050(8)  &   12.86(1)   & 15.46  &  2$\nu_1$   \\ 
		  3  & 13.32457(4)  &    2.74(1)   & 15.08  &  3$\nu_1$   \\ 
		  4  & 17.76622(6)  &    1.89(1)   & 14.91  &  4$\nu_1$   \\ 
		  5  &   7.9972(1)  &    1.05(1)   &  8.35  &  $\nu_2$   \\ 
		  6  &   0.1571(1)  &    0.94(1)   &  7.91  &  $\nu_{\rm low}$   \\ 
		  7  &  22.2078(1)  &    0.80(1)   & 14.30  &  5$\nu_1$   \\ 
		  8  &   5.8212(2)  &    0.58(1)   &  7.75  &  $\nu_3$   \\ 
		  9  &   7.9394(2)  &    0.50(1)   &  5.05  &  $\nu_4$   \\ 
		 10  &   8.0866(2)  &    0.48(1)   &  6.31  &  $\nu_5$   \\ 
		 11  &   7.8392(3)  &    0.40(1)   &  7.76  &  $\nu_6$   \\ 
		 12  &  14.2440(3)  &    0.40(1)   &  7.48  &  $\bf 5\nu_1-\nu_2$   \\ 
		 13  &  26.6492(3)  &    0.39(1)   & 14.04  &  6$\nu_1$   \\ 
		 14  &  11.1479(3)  &    0.39(1)   &  8.40  &  $3\nu_1-\nu_2+\nu_3$   \\
		 15  &  10.2630(3)  &    0.34(1)   &  7.84  &  $\nu_1+\nu_3$   \\
		 16  &  16.9224(3)  &    0.31(1)   &  6.26  &  $\bf2\nu_1+\nu_2$   \\ 
		 17  &  18.6858(3)  &    0.31(1)   &  7.31  &  $\bf6\nu_1-\nu_2$   \\
		 18  &  15.5895(4)  &    0.26(1)   &  8.51  &  $4\nu_1-\nu_2+\nu_3$   \\ 
		 19  &  14.7071(4)  &    0.25(1)   &  6.31  &  $2\nu_1+\nu_3$   \\ 
		 20  &  19.1479(5)  &    0.24(1)   &  7.36  &  $3\nu_1+\nu_3$   \\ 
		 21  &   3.2616(5)  &    0.21(1)   &  7.00  &  $\nu_7$   \\ 
		 22  &  17.0413(5)  &    0.21(1)   &  5.37  &  $\bf 4\nu_1+\nu_3-2\nu_7$   \\ 
		 23  &  31.0909(5)  &    0.20(1)   & 12.69  &  7$\nu_1$   \\ 
		 24  &  23.1276(7)  &    0.16(1)   &  7.13  &  $\bf -2\nu_1+4\nu_2$   \\ 
		 25  &  15.2416(7)  &    0.16(1)   &  5.68  &  $\bf -\nu_1+\nu_2+2\nu_3$   \\ 
		 26  &  21.3646(8)  &    0.14(1)   &  5.13  &  $\bf 3\nu_1+\nu_2$   \\ 
		 27  &  18.5628(9)  &    0.12(1)   &  5.17  &  $5\nu_1+\nu_2-2\nu_3$   \\ 
		 28  &   20.031(1)  &    0.10(1)   &  6.23  &  $5\nu_1-\nu_2+\nu_3$   \\ 
		 29  &   23.589(1)  &    0.10(1)   &  5.68  &  $4\nu_1+\nu_3$   \\ 
		 30  &   21.481(1)  &    0.11(1)   &  5.00  &  $5\nu_1-3\nu_2+4\nu_3$   \\ 
		 31  &   19.684(1)  &    0.09(1)   &  5.40  &  $\bf\nu_2+2\nu_3$   \\ 
		 32  &   35.533(1)  &    0.09(1)   & 10.64  &  8$\nu_1$   \\ 
		 33  &   27.569(2)  &    0.06(1)   &  5.56  &  $-\nu_1+4\nu_2$   \\ 
		 34  &   32.009(3)  &    0.04(1)   &  5.14  &  4$\nu_2$   \\ 
		 35  &   39.974(3)  &    0.04(1)   &  6.39  &  9$\nu_1$   \\ 
		 36  &   36.452(4)  &    0.03(1)   &  5.09  &  $\bf\nu_1+4\nu_2$   \\
    \hline
\enddata
\end{deluxetable}

\begin{deluxetable}{|l|r|r|r|l|}[t]
\tabletypesize{\scriptsize}
			\tablecaption{The frequencies found in the TESS S32 data of RY~Lep. \label{freq_RY_Lep_S32}}
	\tablehead{
                \colhead{ } & \colhead{  Frequency } & \colhead{  A } & \colhead{  S/N } & \colhead{  ID} \\
		  \colhead{ } & \colhead{  $[\rm d^{-1}]$ } & \colhead{  [ppt] } & \colhead{  } & \colhead{}
        }
        \startdata 
		\hline        
		  1  & 4.441567(3)  &   101.24(2)  &  9.47  &  $\nu_1$    \\ 
		  2  &  8.88312(3)  &    12.96(2)  &  9.41  &  $2\nu_1$    \\ 
		  3  &  13.3245(1)  &     2.80(2)  &  9.16  &  $3\nu_1$    \\ 
		  4  &  17.7662(2)  &     1.93(2)  &  9.18  &  $4\nu_1$    \\ 
		  5  &   7.9964(3)  &     1.02(2)  &  5.58  &  $\nu_2$    \\ 
		  6  &   0.1600(4)  &     0.90(2)  &  5.15  &  $\nu_{\rm low}$    \\ 
		  7  &  22.2083(4)  &     0.81(2)  &  9.02  &  $5\nu_1$    \\ 
		  8  &   5.8214(6)  &     0.58(2)  &  5.35  &  $\nu_3$    \\ 
		  9  &  14.2437(9)  &     0.40(2)  &  5.07  &  $5\nu_1-\nu_2$    \\ 
		 10  &  26.6497(9)  &     0.40(2)  &  8.96  &  $6\nu_1$    \\ 
		 11  &   11.149(1)  &     0.34(2)  &  5.36  &  $3\nu_1-\nu_2+\nu_3$    \\ 
		 12  &   10.265(1)  &     0.32(2)  &  5.44  &  $\nu_1+\nu_3$    \\ 
		 13  &   18.686(1)  &     0.31(2)  &  5.17  &  $6\nu_1-\nu_2$     \\ 
		 14  &   31.091(2)  &     0.21(2)  &  8.16  &  $7\nu_1$    \\ 
		 15  &   35.532(4)  &     0.10(2)  &  7.17  &  $8\nu_1$    \\
		\hline
\enddata
\end{deluxetable}
\begin{deluxetable}{|l|r|r|r|l|}[t]
\tabletypesize{\scriptsize}
	\tablecaption{The frequencies found in the TESS S33 data of RY~Lep. \label{freq_RY_Lep_S33}}
	\tablehead{
            \colhead{ } & \colhead{  Frequency } & \colhead{  A } & \colhead{  S/N } & \colhead{  ID} \\
		  \colhead{ } & \colhead{  $[\rm d^{-1}]$ } & \colhead{  [ppt] } & \colhead{  } & \colhead{}
        }
        \startdata 
		\hline
		  1  & 4.441553(3)  &   100.20(2)  & 10.13  &  $\nu_1$    \\ 
		  2  &  8.88309(3)  &    12.78(2)  & 10.03  &  $2\nu_1$    \\ 
		  3  &  13.3245(1)  &     2.67(2)  &  9.65  &  $3\nu_1$    \\ 
		  4  &  17.7662(2)  &     1.85(2)  &  9.55  &  $4\nu_1$    \\ 
		  5  &   7.9995(3)  &     1.07(2)  &  5.68  &  $\nu_2$    \\ 
		  6  &   0.1570(3)  &     0.99(2)  &  5.49  &  $\nu_{\rm low}$    \\ 
		  7  &  22.2076(4)  &     0.80(2)  &  9.13  &  $5\nu_1$    \\ 
		  8  &   5.8196(6)  &     0.60(2)  &  5.01  &  $\nu_3$    \\ 
		  9  &  11.1432(8)  &     0.43(2)  &  5.88  &  $3\nu_1-\nu_2+\nu_3$    \\ 
		 10  &  26.6490(9)  &     0.38(2)  &  8.84  &  $6\nu_1$    \\ 
		 11  &  10.2611(9)  &     0.38(2)  &  5.33  &  $\nu_1+\nu_3$    \\ 
		 12  &  15.5859(1)  &     0.29(2)  &  6.15  &  $4\nu_1-\nu_2+\nu_3$    \\ 
		 13  &   31.091(2)  &     0.20(2)  &  8.13  &  $7\nu_1$    \\ 
		 14  &   35.532(4)  &     0.08(2)  &  6.76  &  $8\nu_1$    \\ 
	\hline
\enddata
\end{deluxetable}

\begin{deluxetable}{|l|r|r|r|l|}
	\tabletypesize{\scriptsize}
	\tablewidth{0pt}
	\tablecaption{The whole set of frequencies found in the ASAS data of RY~Lep, divided into six segments.  \label{freq_RY_Lep_ASAS}}
	\tablehead{
		\colhead{ } & \colhead{  Frequency } & \colhead{  A } & \colhead{  S/N } & \colhead{  ID} \\
		\colhead{ } & \colhead{  $[\rm d^{-1}]$ } & \colhead{  [mmag] } & \colhead{  } & \colhead{}
	}
	\startdata
	\hline \multicolumn{5}{|c|}{\textbf{Total 11.2000 - 12.2009, 3300 d}} \\
	\hline
	1  &  4.441543(3)  &      168(3)  &    17.44  & $\nu_1$\\ 
	2  &  6.59953(3)  &       18(3)  &     4.85  & $\nu_2$\\ 
	3  &  8.88311(2)  &       20(3)  &     4.83  & 2$\nu_1$\\ 
	4  &  6.44450(3)  &       16(3)  &     4.31  & $\nu_1$ + d$^{-1}$?\\ 
	\hline \multicolumn{5}{|c|}{\textbf{11.2000 - 12.2001, 396 d}} \\
	\hline
	1  &  4.44150(4)  &      164(6)  &     8.77  & $\nu_1$ \\ 
	\hline \multicolumn{5}{|c|}{\textbf{07.2002 - 05.2004, 680 d}} \\
	\hline
	1  &  4.44157(2)  &      171(5)  &    10.24  &  $\nu_1$\\ 
	2  &  6.5998(1)  &    29(5) &     5.51  &  $\nu_2$\\ 
	3  &  10.8862(2)  &       21(5)  &     4.38  & 2$\nu_1$+2\,d$^{-1}$?\\ 
	\hline \multicolumn{5}{|c|}{\textbf{09.2004 - 05.2006, 601 d}} \\
	\hline
	1  &  4.44164(4)  &      162(8)  &     6.94  & $\nu_1$ \\ 
	\hline \multicolumn{5}{|c|}{\textbf{12.2006 - 05.2008, 513 d}} \\
	\hline
	1  &  4.44159(3)  &      172(5)  &     8.00  & $\nu_1$ \\ 
	\hline \multicolumn{5}{|c|}{\textbf{08.2008 - 12.2009, 483 d}} \\
	\hline
	1  &  4.44146(4)  &      159(4)  &     6.21  & $\nu_1$ \\ 
	2  &  8.6025(2)  &       28(4)  &     4.62  & $\nu_2$+2\,d$^{-1}$? \\ 
	\hline
	\enddata
\end{deluxetable}

\begin{deluxetable}{|l|r|r|r|l|}
	\tablecaption{The whole set of frequencies found in the SuperWASP data of RY~Lep, divided into three segments.  \label{freq_RY_Lep_superwasp}}
	\tablehead{
		\colhead{ } & \colhead{  Frequency } & \colhead{  A } & \colhead{  S/N } & \colhead{  ID} \\
		\colhead{ } & \colhead{  $[\rm d^{-1}]$ } & \colhead{  [mmag] } & \colhead{  } & \colhead{}
	}
	\startdata          
	\hline \multicolumn{5}{|c|}{\textbf{10.2007 - 03.2008, 152 d}} \\
	\hline
	1  &  4.44154(4) &      155(2)  &     8.58  & $\nu_1$ \\ 
	2  &  8.8831(3)  &       23(2)  &     5.88  & $2\nu_1$ \\ 
	3  &  6.5994(3)  &       21(2)  &     6.99  & $\nu_2$\\ 
	4  &  2.0015(2)  &       40(2)  &     4.90  & ? \\ 
	5  &  5.0045(5)  &       15(2)  &     5.21  & ?\\ 
	6  &  10.0297(7)  &       10(2)  &     4.48  & ?\\
	\hline \multicolumn{5}{|c|}{\textbf{09.2009 - 03.2010, 158 d}} \\
	\hline
	1  &  4.4420(3)  &      123(8)  &     5.68  &$\nu_1$ \\ 
	\hline \multicolumn{5}{|c|}{\textbf{09.2011 - 03.2012, 153 d}} \\
	\hline
	1  &  4.44161(4) &      147(1)  &    12.91  & $\nu_1$\\ 
	2  &  6.6005(1)  &       51(1)  &    11.43  & $\nu_2$\\ 
	3  &  11.0424(3)  &       23(1)  &    10.15  & $\nu_1+\nu_2$\\ 
	4  &  8.8835(3)  &       18(1)  &     9.25  & $2\nu_1$\\ 
	5  &  2.1585(4)  &       13(1)  &     4.63  & $\nu_2-\nu_1$\\ 
	6  &  0.9993(4)  &       19(1)  &     4.35  & ? \\ 
	7  &  8.087(1)   &        6(1)  &     4.59  & $\nu_3$ \\ 
	8  &  15.484(1)   &        5(1)  &     4.58  & $2\nu_1+\nu_2$\\ 
	9  &  17.640(2)   &        4(1)  &     4.05  & $\nu_1+2\nu_2$\\ 
	\hline
	\enddata
\end{deluxetable}

\restartappendixnumbering

\section{The histograms from seismic modeling with Bayesian analysis}\label{sec:appendix_histograms}

\begin{figure*}
    \centering
    \includegraphics[width=0.9\textwidth]{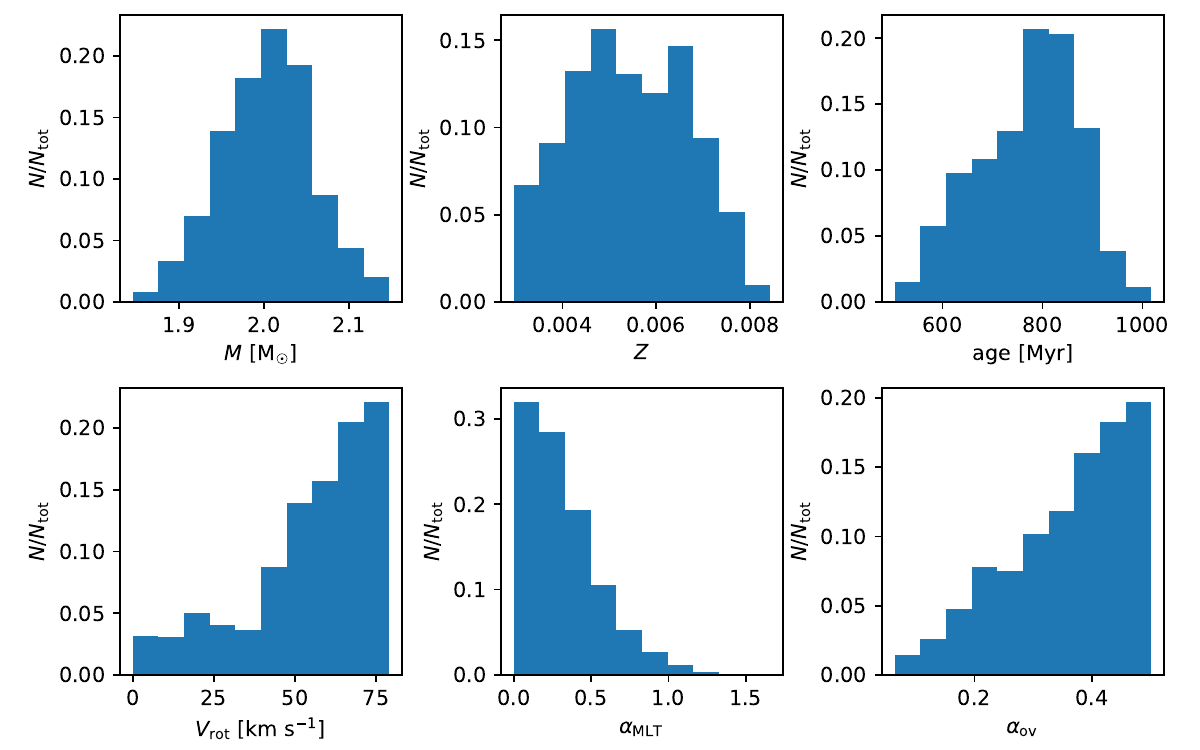}
    \caption{The histogram for parameters obtained from complex seismic modeling of the star RY~Lep. Models reproduce the frequencies $\nu_1=4.441626$\,d$^{-1}$ and $\nu_2=6.5991$\,d$^{-1}$ as first and third overtone radial modes, respectively. The microturbulency velocity $\xi = 4$\,km\,s$^{-1}$ and the initial hydrogen abundance $X_0 = 0.7$ are adopted.}%
    \label{fig:rylep_l0_l0_histogram}%
\end{figure*}

This section presents histograms of the parameters of the seismic models of RY~Lep obtained from Monte Carlo simulations combined with Bayesian inference. In Fig.~\ref{fig:rylep_l0_l0_histogram}, we show histograms for the model parameters derived under the assumption that the frequencies near $\sim 4.44$\,d$^{-1}$ and $\sim 6.6$\,d$^{-1}$ correspond to two radial modes$-$the first and third overtones, respectively. The plotted parameters include the mass $M$, metallicity $Z$, age, current rotational velocity $V_{\rm rot}$, mixing-length parameter $\alpha_{\rm MLT}$, and overshooting parameter $\alpha_{\rm ov}$.

\begin{figure*}
    \centering
    \includegraphics[width=0.9\textwidth]{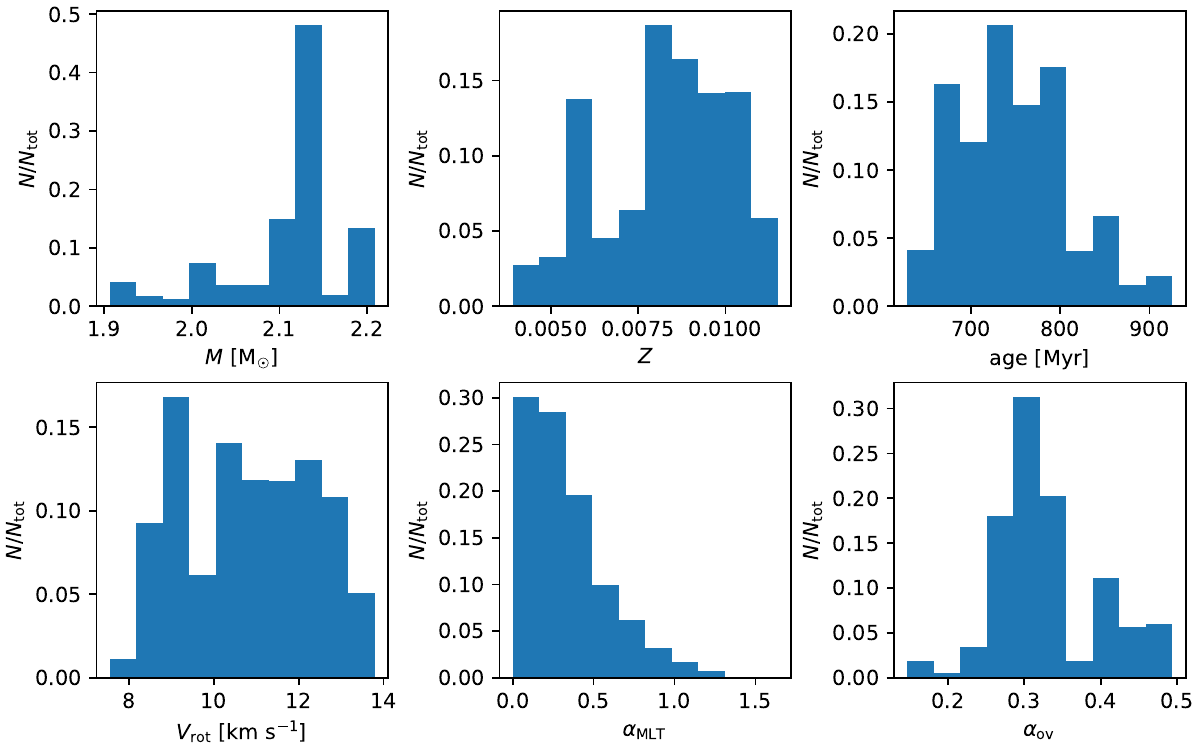}
    \caption{The same as in Fig.~\ref{fig:rylep_l0_l0_histogram} with $\nu_2=6.5991$\,d$^{-1}$ as a nonradial mode ($\ell=1$, $m=-1$).}%
    \label{fig:rylep_l0_l1_mm1_histogram}%
\end{figure*}

\begin{figure*}
    \centering
    \includegraphics[width=0.9\textwidth]{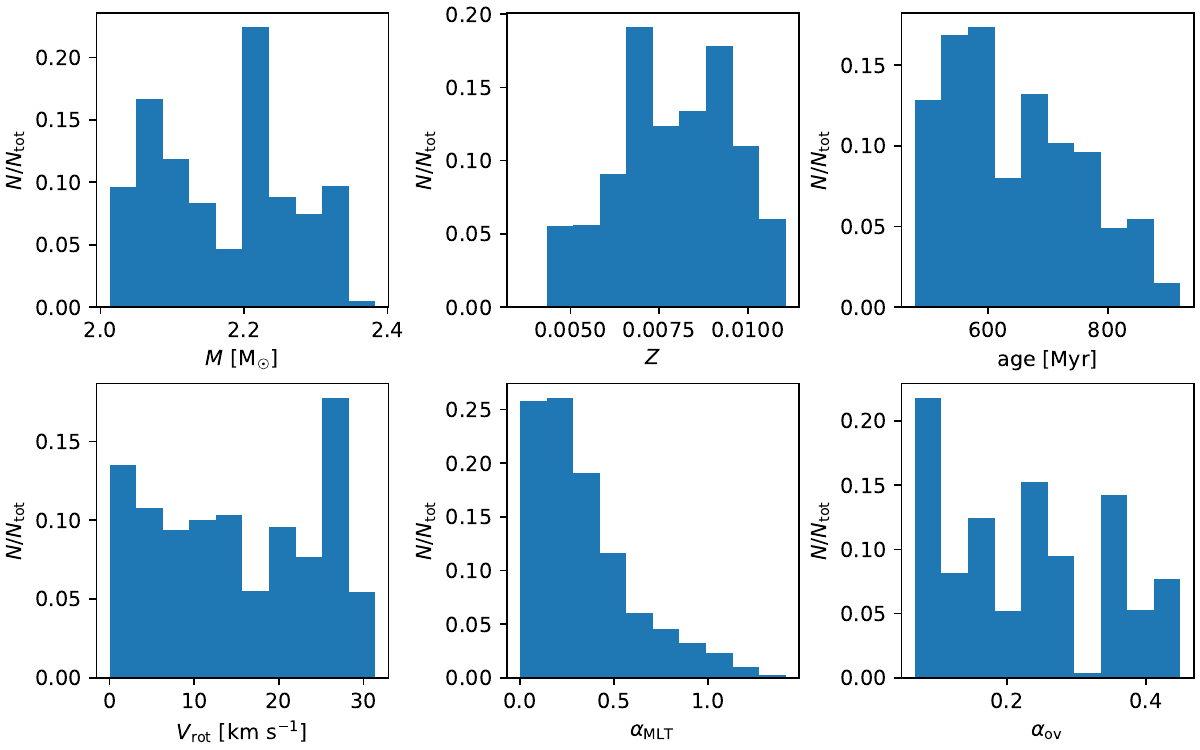}
    \caption{The same as in Fig.~\ref{fig:rylep_l0_l0_histogram} with $\nu_2=6.5991$\,d$^{-1}$ as a nonradial mode ($\ell=1$, $m=0$).}%
    \label{fig:rylep_l0_l1_m0_histogram}%
\end{figure*}

\begin{figure*}
    \centering
    \includegraphics[width=0.9\textwidth]{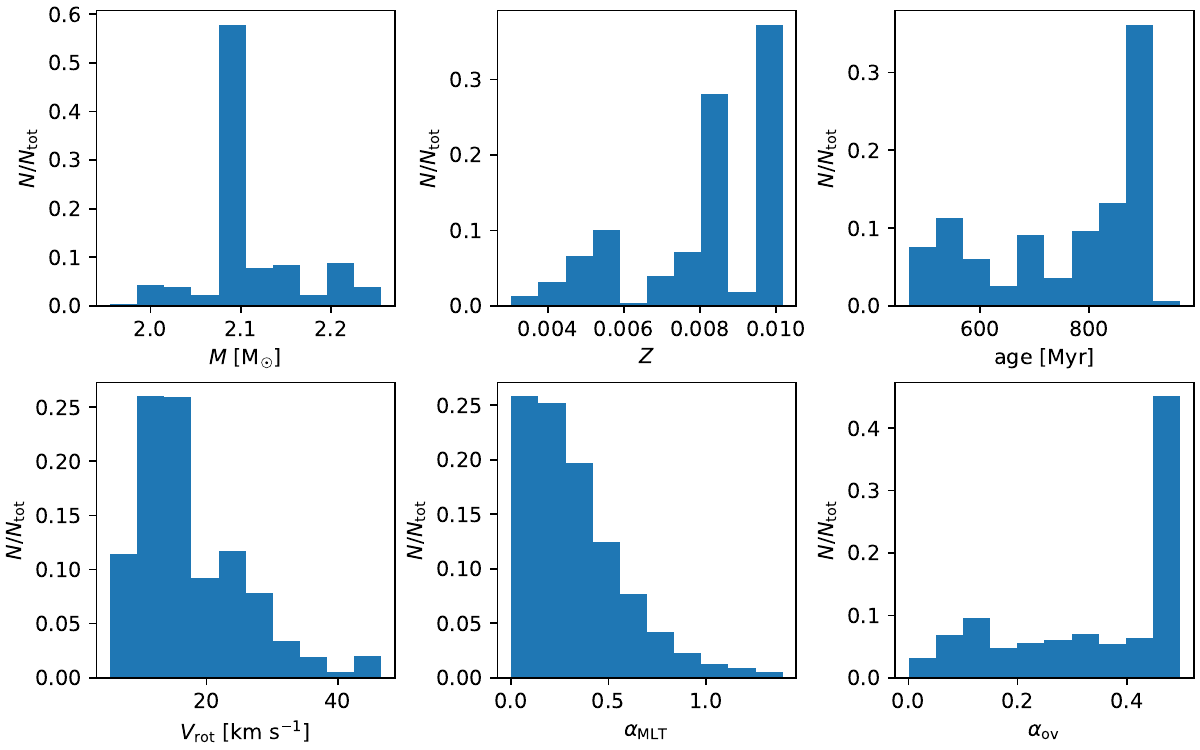}
    \caption{The same as in Fig.~\ref{fig:rylep_l0_l0_histogram} with $\nu_2=6.5991$\,d$^{-1}$ as a nonradial mode ($\ell=1$, $m=1$).}%
    \label{fig:rylep_l0_l1_mp1_histogram}%
\end{figure*}

Figures~\ref{fig:rylep_l0_l1_mm1_histogram}–\ref{fig:rylep_l0_l1_mp1_histogram} display histograms for the same set of parameters, but computed under the alternative hypothesis that the $\sim4.44$\,d$^{-1}$ frequency corresponds to the first radial overtone and the $\sim 6.6$\,d$^{-1}$ frequency corresponds to a dipole ($\ell = 1$) mode with $m = -1, 0,$ or $+1$.

\clearpage

\bibliography{biblio}{}
\bibliographystyle{aasjournalv7}



\end{document}